\documentclass[a4paper,superscriptaddress,floatfix,twocolumn,nofootinbib, longbibliography, notitlepage]{revtex4-2}
\usepackage{bbm}
\usepackage{bbold}
\usepackage{bbm}
\usepackage{algpseudocode}
\usepackage[pdftex]{graphicx}
\usepackage{latexsym,amsmath,verbatim,amssymb,txfonts}
\usepackage{lipsum}
\usepackage{color}
\usepackage{rotating}
\usepackage{comment}
\usepackage{verbatim}
\usepackage{physics}
\usepackage{braket}
\usepackage{multirow}
\usepackage{hyperref}
\usepackage[english]{babel}
\usepackage{bm}
\usepackage{amsmath}
\usepackage{amsfonts}
\usepackage{amssymb}
\usepackage{float}
\usepackage{color}
\usepackage{braket}
\usepackage{ulem}
\usepackage{blkarray, bigstrut}

\renewcommand\bra[1]{{\left\langle{#1}\right|}}
\renewcommand\ket[1]{{\left|{#1}\right\rangle}}
\newcommand\mean[1]{{\left\langle{#1}\right\rangle}}
\def\be{\begin{eqnarray}}
\def\ee{\end{eqnarray}}

\usepackage{dsfont} 
\newcommand{\idop}{\mathds{1}}

\begin{document}
\title{Weak continuous measurements require more work than strong ones}
\author{Lorena Ballesteros Ferraz}
\email{lorena.ballesteros-ferraz@cyu.fr}
\affiliation{Université de Lorraine, CNRS, LPCT, F-54000 Nancy, France}
\affiliation{Laboratoire de Physique Théorique et Modélisation, CNRS Unité 8089,
CY Cergy Paris Université, 95302 Cergy-Pontoise cedex, France}
\author{Cyril Elouard}
\email{cyril.elouard@univ-lorraine.fr}
\affiliation{Université de Lorraine, CNRS, LPCT, F-54000 Nancy, France}

\begin{abstract}
Understanding the energy cost of quantum measurement process and its connection to the measurement performance faces the challenge of modeling the objectification process. The latter, turns the measurement result into an objective fact, available to independent observers, and is responsible for the measurement irreversibility. To address this issue, we propose and analyze a dynamical model of quantum measurement, able to capture nonideal (weak and inefficient) measurements. In this model, the objectification is induced by a contact with a macroscopic reservoir at equilibrium which is responsible for the redundant broadcast of the measurement outcome (producing a Spectrum Broadcast Structure (SBS) state) while inducing decoherence in the pointer basis, in the line of the theory of quantum Darwinism.
We analyze the performance of the obtained measurement process by introducing figures of merit to quantify the strength of the measurement and its efficiency. 
We also derive and a lower bound on the measurement work cost that we can relate to the measurement quality. We take as an illustration the readout of a qubit via its coupling to a harmonic oscillator. We investigate the long sequences of extremely short and weak measurements (a.k.a continuous measurements), to find under which conditions they converge to an ideal (projective) measurement and analyze their work cost. Surprisingly, we find that a sequence converging to projective measurement has a much larger work cost than an equivalent strong measurement obtained from a single intense interaction with the apparatus. We extend this result to a large class of models owing to scaling arguments. Our analysis offers new insights into the trade-offs between measurement strength, energy consumption, and information extraction in quantum measurement protocols.
\end{abstract}

\maketitle

\section{Introduction}
Quantum measurements are among the most intriguing processes in quantum mechanics, and their modeling has been the subject of extensive discussion for decades~\cite{braginsky1995quantum, Allahverdyan13,brukner2017quantum, griffiths2017quantum}. Quantum theory allows for various types of measurements ranging from the ideal projective measurements - introduced by the measurement postulate of quantum mechanics - to more sophisticated processes, such as POVMs (Positive Operator-Valued Measures), quantum non-demolition measurements, and homodyne detection~\cite{busch2016quantum, wiseman2009quantum,Jordan242}. In practical experiments, measurement protocols are non-ideal due to factors such as detector inefficiency, thermal noise, imperfections in state preparation and finite interaction time with the measured system. These sources of non-ideality can significantly impact the measurement process, and their effects can be modeled and quantified~\cite{auffeves2019generic, warszawski2002quantum, wiseman2009quantum}.

All quantum measurement processes share the fundamental requirement to generate a family of macroscopically distinguishable states of the apparatus, that can be read owing to conventional (non quantum) apparatuses, yielding a measurement result (or outcome). The latter is a classical stochastic variable whose value can be stored into a classical memory device. The transition towards a classical behavior of the degrees of freedom of the apparatus encoding the measurement result inherently involves a non-unitary process~\cite{neumann1955mathematical, joos2013decoherence}. It can therefore be derived from a purely unitary evolution of the system and the apparatus only if the latter contains many degrees of freedom, in addition to the ones accessible to the observer, over which a trace can be taken.
This requirement is closely related to the theory of quantum Darwinism, which posits that the environment facilitates the emergence of classicality by encoding redundant copies of information (i.e. of the measurement result), thereby making it accessible to multiple observers and hence an objective fact~\cite{zurek2009quantum}. 

Since quantum measurements cannot be modeled by a purely unitary process of the system and a pointer, they constitute a non-trivial thermodynamic transformation, involving exchanges of entropy and energy. These thermodynamic resources have been extensively explored, notably in the context of powering quantum engines and cooling mechanisms~\cite{parrondo2015thermodynamics, elouard2018efficient, buffoni2019quantum,Jordan20,Fadler23}. Despite the central role of quantum measurements in virtually all quantum technologies and applications~\cite{fellous2023optimizing, pirandola2020advances, degen2017quantum}, relatively few studies have addressed the work cost associated with performing quantum measurements~\cite{Jacobs09,Sagawa2009Jun,deffner2016quantum, debarba2019work,Mohammady2023Jan,Piccione2023Nov,latune2024thermodynamically}. A key result found in the literature is that ideal projective measurements, in principle, require infinite resources, making it impossible to achieve them exactly. Instead, they can only be (arbitrarily well) approximated~\cite{guryanova2020ideal}.

One might ask what the minimum amount of work required for a quantum measurement is and what the optimal protocol would be to achieve this minimum. In a previous study, some of us derived a lower bound on the work cost required to perform a quantum measurement in terms of figures of merit quantifying the information acquired by the measurement \cite{latune2024thermodynamically}. To obtain such a result, the measuring apparatus was modeled as a pointer continuously coupled to a reservoir able to bring it back to thermal equilibrium. The obtained lower bound was then based on the expression of the Second Law of thermodynamics during the measurement process. Deviations from the minimum work cost were associated with the entropy produced when operating and resetting the measurement device. Surprisingly, a thermodynamically reversible measurement protocol, achieving asymptotically the minimum cost, was identified and analyzed. However, this setup could not capture the paradigmatic case of a weak (incomplete) but efficient (purity-preserving) measurement~\cite{jacobs2006straightforward}.

In this work, we investigate two new questions: 
First, is it optimal to achieve classicality owing to a thermalizing bath (in which energy can be wasted under the form of heat dissipation), or is a pure dephasing process (comprising no heat transfer) preferable? Surprisingly, we find that a pure dephasing bath actually leads to a less favorable lower bound on the work cost.
Second, how is the energy cost scaling with the strength of the measurement (the amount of extracted information): In other words, is it more energetically favorable to perform a sequence of weak measurements converging towards a projective measurement~\cite{Jacobs2006Sep}, or a single nearly-projective measurement? Our analysis shows that, despite the work cost of the measurement vanishing with the measurement strength, the converging sequence of weak measurements requires much more energy in general, revealing a previously unrecognized fundamental trade-off, which constitutes the central result of this work.

The paper is organized as follows: First, we introduce in Sec.~\ref{section:modelling_the_dephasing_protocol} the quantum measurement protocol considered throughout the paper. Next, we examine in Sec.~\ref{section:energetics} the energetics of the dephasing process and derive the lower bound on the measurement work cost. Sec.~\ref{section:qubit-cavity-model} focuses on a specific model involving a qubit and a harmonic oscillator (cavity). In Sec.~\eqref{section:strength_and_efficiency}, we introduce and discuss measures of non-ideality, specifically strength and efficiency of the measurement. We use them in Sec.~\ref{section:single_measurement} to analyze the work cost and the performance of a single qubit measurement, and in Sec.~\ref{section:measurement_sequence} to compare long sequences of weak continuous measurements with single strong measurement.

\begin{figure} [h!]
    \centering
\includegraphics[width=0.8
\linewidth]{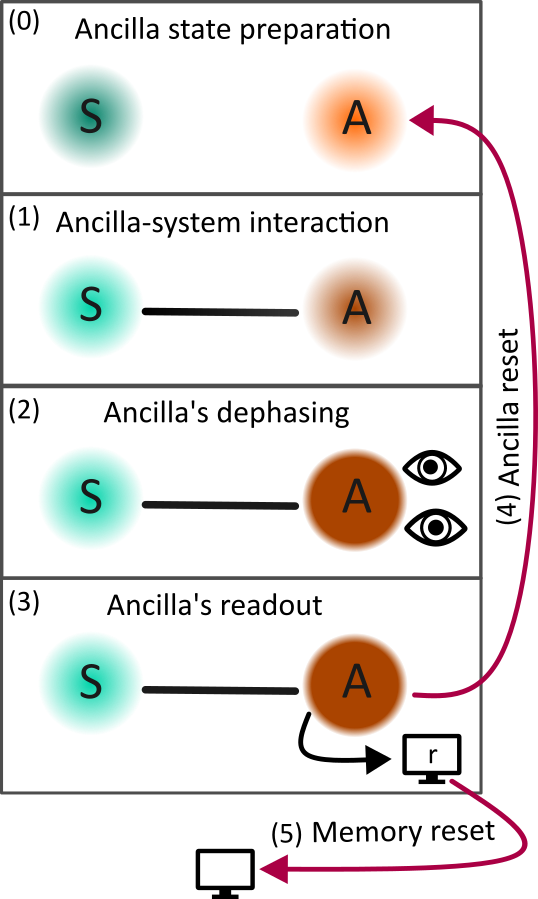}
    \caption{Measurement protocol: (0) ancilla state preparation, (1) ancilla-system interaction, (2) ancilla dephasing, (3) ancilla readout, (4) ancilla reset and (5) memory reset. The color change in both the system and ancilla indicates a change of state. Blurred disks represent quantum states, while filled disks denote classical states.}
    \label{fig:measurement-protocol}
\end{figure}

\section{Dynamical model of quantum measurement} \label{section:modelling_the_dephasing_protocol}
In this article, we consider the following measurement protocol involving a system S in state $\hat{\rho}_S$ (the measured system) and an ancilla A, initialized at time $t_0$ in state $\hat{\rho}_A(t_0)$ (Step (0) in Fig.~\ref{fig:measurement-protocol}). The ancilla A models both the part of the apparatus that first interacts with S, as well as the degrees of freedom in which the measurement result will be read by the observer. In contrast, the influence of other inaccessible degrees of freedom of the apparatus will be included via the non-unitary dynamics they induce on A. The combined system-ancilla state $\hat{\rho}_{SA}\left(t_0\right) = \hat{\rho}_S(t_0) \otimes \hat{\rho}_A(t_0)$ is initially separable, meaning no information about S is {\it a priori} contained in A. Next, the system and the ancilla unitarily interact, transferring information from the system to the ancilla (step (1) in Fig.~\ref{fig:measurement-protocol}). The interaction lasts for $\Delta t = t_1 - t_0$ and is governed by the unitary operator $\hat{U}(\Delta t)$, which possibly also includes control operations affecting only the ancilla. The full system-ancilla state after the interaction is
\begin{eqnarray}
\hat{\rho}_{SA}\left(t_1\right)&=&\hat{U}\left(\Delta t\right)\left(\hat{\rho}_S\left(t_0\right)\otimes\hat{\rho}_A\left(t_0\right)\right)\hat{U}^{\dagger}\left(\Delta t\right).
\end{eqnarray}

The amount of correlations that are generated between the system and the ancilla (or equivalently, the amount of information about the system extracted during the interaction) is set by the coupling constant and the interaction time $\Delta t$: in the case of a weak interaction, or short interaction time, only a small amount of information is transferred, and probing the ancilla will only weakly affect the system state. Conversely, a strong/long enough interaction time may fully copy in the ancilla all the information about the system observable involved in the interaction (that is, generate maximal correlations between the ancilla and the system states), leaving the system in a fully mixed reduced state~\cite{jacobs2006straightforward}. 

Following the interaction step, we assume that the ancilla undergoes a strong pure dephasing process (step (2) in Fig.~\ref{fig:measurement-protocol}). This step models the ``objectification process'', during which the information about the system, stored in the ancilla, is redundantly copied multiple times into independent subsets of macroscopic amount of degrees of freedom, thereby, making the measurement result an objective fact~\cite{zurek2009quantum,Mohammady2021Dec,latune2024thermodynamically}. 

\begin{figure} [h!]
    \centering
\includegraphics[width=0.8
\linewidth]{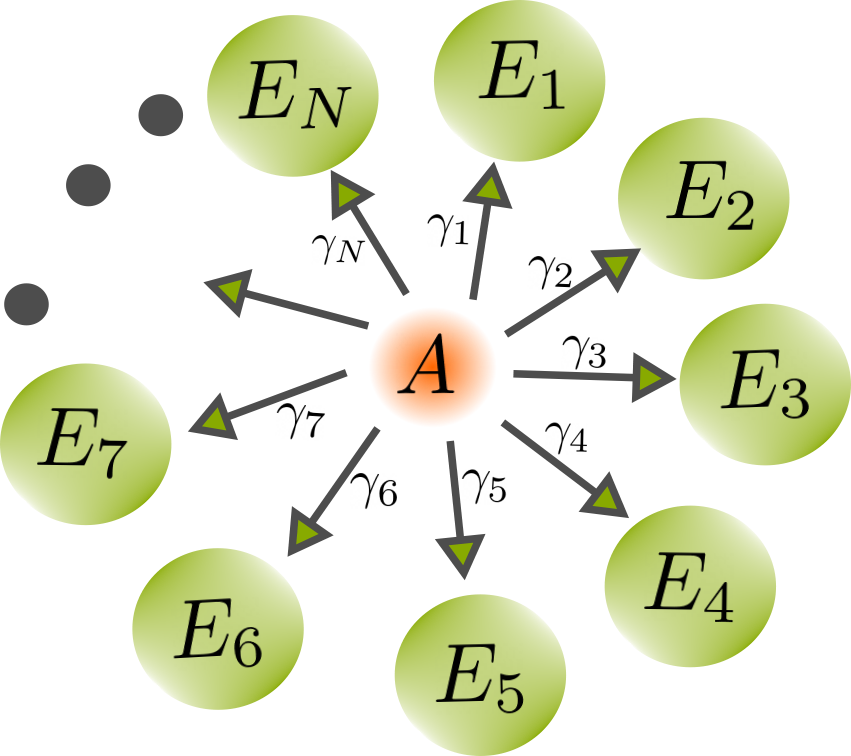}
\caption{\label{fig:dephasing_different_channels}The ancilla undergoes dephasing through multiple independent environments. This occurs via distinct channels, each characterized by a dephasing coefficient $\gamma_i$. At the end of the process, the joint state of the ancilla and the complete system is in a Spectrum Broadcast Structure (SBS) state.}
\end{figure}

This process can arise from internal dynamics of a macroscopic system~\cite{Engineer2024Mar} - the inaccessible degrees of freedom of the apparatus - or from the irreversible processes taking place when the signal from the apparatus is extracted and amplified. The objectification process selects a basis of stable (pointer) states of the ancilla associated to the different measurement outcomes, while coherences between those pointer states are completely suppressed. 

To model this step, we assume that the ancilla interacts simultaneously with distinct, non-interacting subsets of the macroscopic degrees of freedom (or, in the language of open quantum systems, separate environments), with potentially different dephasing interaction coefficients, as illustrated in Fig.~\ref{fig:dephasing_different_channels}. Once dephasing has lasted long enough to fully decohere the ancilla state, the joint state of the ancilla and any of the sub-environments takes the form of a so-called SBS state~\cite{horodecki2015quantum}, which ensures that the measurement outcome could be retrieved independently from a measurement in each of the separate sub-environments (See Appendix~\ref{appendix:SBS_formation} for derivation and further details).

Pure dephasing, that is, decoherence in the absence of energy exchange with the environment, naturally emerges from such microscopic models whenever the ancilla-reservoir coupling commutes with the ancilla's Hamiltonian. The resulting dynamics then show the damping of coherences in the ancilla’s energy eigenbasis $\{\ket{n}\}$, which therefore plays the role of a pointer basis. Pure-dephasing in another basis can be achieved in engineered environments resulting from driven-dissipative processes (e.g. in a quantum limited amplifier~\cite{Flurin2014Thesis}). Those processes amount to temporarily modify the effective Hamiltionian of the ancilla (e.g. owing to an external driving), such that the interaction selects another pointer basis.

Furthermore, we require that this objectification process encompasses any evolution with nontrivial energetic and thermodynamic role involved in the measurement process, by imposing that at the end of step (2), the ancilla is fully dephased, i.e. in a classical mixture of the pointer states:

\begin{eqnarray}
\label{eq:full-system-ancilla-state-after-dephasing}
\hat{\rho}_{SA}\left(t_f\right)&=&\sum_n\ket{n}\bra{n}\hat{U}\left(\Delta t\right)\left(\hat{\rho}_S\left(t_0\right)\otimes\hat{\rho}_A\left(t_0\right)\right)\hat{U}^{\dagger}\left(\Delta t\right)\ket{n}\bra{n}\nonumber\\ 
&=&\sum_n p_n\hat{\rho}_{S|n}\otimes\ket{n}\bra{n},\\
   p_n=\text{Tr}&\Big[\Big(&\hat{I}\otimes\hat{\Pi}_n\Big)\hat{U}\left(\Delta t\right)\left(\hat{\rho}_S\left(t_0\right)\otimes\hat{\rho}_A\left(t_0\right)\right)\hat{U}^{\dagger}\left(\Delta t\right)\left(\hat{I}\otimes\hat{\Pi}_n\right) \Big].\quad\;\nonumber
\end{eqnarray}
where $\hat{\rho}_{S|n}$ (resp. $p_n$) is the conditional system state associated with (the probability) finding the ancilla in the state $\ket{n}$.

Finally, the ancilla is read out in the basis set by the dephasing process, here the energetic eigenbasis ((3) in Fig.~\ref{fig:measurement-protocol}). We assume that the dephasing process singles out a set of macroscopically distinguishable states of the ancilla, which can be read via a classical measurement (either directly in the ancilla if it is a macroscopic degree of freedom, like a large-amplitude field mode, or indirectly via its impact in the environment, for instance via the light it emits). As we required the dephasing to be complete during the objectification step, this process corresponds only to information acquisition, without exchange of energy. 

The only impact of the readout process in the quantum state describing the system and the ancilla is the selection of an element of the classical mixture, just as a measurement performed on a probabilistic ensemble of classical systems. While in principle all $\ket{n}$ can be perfectly discriminated (being orthogonal quantum states), it is more reasonable to assume that macroscopically distinguishable observables will correspond to coarse-grained subsets of those states. A convenient way to introduce such subsets is via a set of orthogonal projectors $\hat{\Pi}_r = \sum_n p(r|n) \hat \Pi_n$, where $\hat\Pi_n = \ket{n}\bra{n}$ and where the index $r$, the measurement outcome, labels the distinguishable coarse-grained subspaces. We furthermore require $\sum_r \hat\Pi_r = \idop$ and $\hat\Pi_r \hat\Pi_{r'} = \delta_{r,r'}\hat\Pi_r$ such that $p(r|n)$ is a piecewise-constant function with values in $\{0,1\}$. Injecting the latter relation  into Eq.~\eqref{eq:full-system-ancilla-state-after-dephasing} leads to:

\be\label{eq:av_state_SA}
\hat{\rho}_{SA}\left(t_f\right)&=& \sum_r p_r {\hat{\rho}_{SA|r}}(t_f),
\ee
with 
\be\label{eq:definition_rho_SA_given_r}
{\hat{\rho}_{SA|r}}(t_f) = \frac{1}{p_r}\sum_n p_n p(r|n) {\hat{\rho}}_{S|n}\otimes\hat\Pi_n,
\ee
and $p_r = \sum_n p_n p(r|n)$ the probability of obtaining the outcome $r$.

After reading the ancilla's state and storing the information in a classical memory, the measurement process is practically over. However, to make sure all thermodynamic costs are included in our protocol, we also include the reset of the ancilla and the memory (after its information content has been potentially used for some task) to their initial states ((4 \& 5) in Fig.~\ref{fig:measurement-protocol}.

This measurement protocol can model a wide range of quantum measurements, from strong to weak and from ideal to noisy. Like the protocol analyzed in~\cite{latune2024thermodynamically}, it does not involve any black-box quantum measurement on the ancilla which would make the thermodynamic analysis incomplete.

\section{Thermodynamic analysis}\label{section:energetics}
In this section, we analyze step-by-step the energetic cost of the protocol presented in section~\ref{section:modelling_the_dephasing_protocol}, and the bounds the Second law of thermodynamics imposes on it. 

(0). \underline{Ancilla state preparation:}\\
We assume that the ancilla is at time $t_0$ in a free state at equilibrium with its environment at inverse temperature $\beta$.

(1). \underline{System-ancilla interaction:}\\
In this step, neither the system nor the ancilla interact with the environment, so no heat is exchanged. However, work may be exchanged with external drives to control the coupling between the system and the ancilla and to drive the ancilla directly:
\begin{equation}\label{eq:Wdr}
    \text{W}_{\text{dr}}= E_{SA}(t_1)-E_{SA}(t_0)= \Delta E_S + E_A(t_1)-E_A(t_0),
\end{equation}
where $E_i(t)=\Tr\{\hat{H}_i\hat\rho_i(t)\}$ represents the energy of system $i=S,A$ at time $t$, with $\hat\rho_i(t)$ the reduced state of system $i$. Above, $t_1$ is the final time of step $1$. This energy exchange arises from the dynamical evolution of the full-system state governed by the system-ancilla Hamiltonian. In Eq.~\eqref{eq:Wdr}, we have assumed that the coupling between the ancilla and the system vanishes at $t_0$ and at times $t>t_1$, and that $\hat H_A(t_0)=\hat H_A(t_1)=\hat H_A$. 

Moreover, the pure dephasing process (step~2) only acts on the ancilla, leaving the system's energy unaffected. Consequently, the system energy satisfies $E_S(t_1)-E_S(t_0)=E_S(t_f)-E_S(0)\equiv\Delta E_S$, where $t_\text{f}$ denotes the time of the ancilla readout at the end of step~3. The system's energy remains unchanged throughout the rest of the protocol.

(2 \& 3). \underline{Dephasing on the ancilla and readout:}\\
We first analyze the case where the pure dephasing occurs in the eigenbasis of $\hat H_A$. During the dephasing process, there is no exchange of energy — neither work nor heat — with the systems. In the strong coupling regime, a nonzero work cost may be associated to the action of coupling and decoupling the system from the environment~\cite{mitchison}. However, the latter is not fundamental as it vanishes in the case where the coupling Hamiltonian to the environment commutes with the environment's Hamiltonian (see Appendix~\ref{appendix:SBS_formation}.). This is verified e.g. in the dispersive regime where the environment modes are far from resonance with the ancilla. We therefore exclude this cost from our lower bound. The second law then simply takes the form:

\begin{equation}
    \sigma=\Delta S_{SAM}\geq 0, 
\end{equation}
where $\sigma$  denotes the entropy produced during the steps $(2 \& 3)$ and $\Delta S_{SAM}$ the change in von Neumann entropy of the system–ancilla–memory composite between the initial state and the state after the dephasing procedure. Additionally, the classical readout of the ancilla can, in principle, be implemented at no work cost \cite{landauer1961irreversibility}. 

To provide a bound for the case where pure dephasing occurs in another basis, we consider that the Hamiltonian of the ancilla is switched to another Hamiltonian $\tilde H_A$ before the pure dephasing process starts. The latter is then assumed to damp coherences in the eigenbasis of $\tilde H_A$. Finally, the Hamilltonian of $A$ is switched back to $H_A$. Assuming instantaneous quench-like variations of the Hamiltonian, we identify an additional work cost~\cite{Esposito10}:
\begin{eqnarray}
    W_\text{sw} &=& \text{Tr}\{(\tilde H_A-H_A)(\hat\rho_A(t_1)-\hat\rho_A(t_f)\}\nonumber\\
    &=& E_A(t_1)-E_A(t_f).
\end{eqnarray}
In the last line, we have used that the pure dephasing process conserves the expectation value of $\tilde H_A$. As before, the energy of $A$ is defined with respect to Hamiltonian $H_A$.

(4 \& 5). \underline{Reset of the ancilla and the classical memory:} \\
Finally, first the ancilla and then the classical memory must be reset to their initial states. The minimum work required for the ancilla reset in an environment of inverse temperature $\beta$ is
\begin{equation}
    W^A_{\text{reset}}\geq E_A(t_0)-E_A(t_f)+\frac{1}{\beta}\Delta \mean{S_A},
\end{equation}
where 
\be
\Delta \mean{S_A}=\sum_r p_r\left(S[\hat{\rho}_{A|r}(t_f)]-S_A\left(t_0\right)\right),
\ee
represents the average change in entropy of the ancilla, realization-wise, expressed in terms of $\hat\rho_{A|r}=\text{Tr}_S\{\hat\rho_{SA|r}\}$. Based on the block-diagonal structure of $\hat{\rho}_{SA|r}$ in Eq.~\ref{eq:definition_rho_SA_given_r}, the average entropy of the conditional ancilla states fulfills \cite{nielsen2010quantum}:
\begin{equation}
    \sum_r p_rS[\hat{\rho}_{A|r}(t_f)]=H\left(p_n\right)+\sum_n p_n H\left(p_{r|n}\right)-H\left(p_r\right),
\end{equation}
where $H\left(p_i\right)$ denotes the Shanon entropy of the probability distribution $p_i$.

The term $\sum_n p_n H\left(p_{r|n}\right)$ vanishes whenever the conditional ancilla states $\hat{\rho}_{A|r}$ are orthogonal, which is true here as the coarse-graining projectors $\hat{\Pi}_r$ are themselves orthogonal. Therefore, $\Delta \mean{S_A} = H(p_n)-H(p_r)-S_A\left(t_0\right)$. 

On the other hand, Landauer's principle states that the work required to reset the classical memory fulfills~\cite{landauer1961irreversibility}:
\begin{equation}
    W_{\text{reset}}^M\geq \frac{1}{\beta}H\left(p_r\right).
\end{equation}

Finally, the minimum work required to complete the entire measurement protocol, takes the simple form
\begin{equation}
\label{eq:work_bound_dephasing}
W_{\text{dr}}+W_{\text{reset}}\geq \frac{1}{\beta}(H\left(p_n\right)-S_A(t_0))+\Delta E_S,
\end{equation}
which is independent on the choice of coarse-graining (provided the coarse-graining spaces are orthogonal), and of the pure-dephasing basis. 

If we start with a pure initial ancilla state, it is straightforward to see that the work bound is bounded by $\log d_A$, where $d_A$ is the dimension of the Hilbert space of $A$. Therefore, a lower value is to be expected for small ancilla systems such as qubits than for higher-dimensional systems like a harmonic oscillator. However, we recall that in our description, the ancilla must be able to achieve macroscopically distinguishable pointer states, as does the amplified output of a measuring apparatus. It must therefore be a many-level system that is read out in a coarse-grained fashion (hence a typically small number of different outcomes). In the next section, we model the pointer as a harmonic oscillator so as to explore the consequences of this coarse-graining on the tradeoff between measurement cost and efficiency.

In the case where the bath is dissipative rather than solely causing dephasing, the work bound is given by~\cite{latune2024thermodynamically} (see Apendix~\ref{appendix:work_bound_dissipation})\\
\begin{eqnarray}
\label{eq:work_dissipation_main}
  &&W^{\text{dissipation}}=W_{\text{dr}}+ W^{A}_{\text{reset}}+ W^{M}_{\text{reset}} \\ \nonumber
  &&  \geq \Delta E_{S} +\frac{1}{\beta}(H\left(p_n\right)-S_A(t_0)) -\frac{1}{\beta} \Delta S_{SA},
\end{eqnarray}
where $\Delta S_{SA}$ is the variation of the system and ancilla entropy due to the dephasing process, which is necessary nonnegative, as entropy can only increase or remain constant during unread measurements~\cite{nielsen2010quantum}.
The formal difference between the work bounds in Eq.~\ref{eq:work_bound_dephasing} and Eq.~\ref{eq:work_dissipation_main} is therefore $ -\frac{1}{\beta} \Delta S_{SA}$, such that the dissipation work bound is always less than or equal to the dephasing work bound. This is a first result of our analysis, which goes against the intuition that heat dissipation in the measuring apparatus is detrimental to the energy cost. The reason for that is the possibility in the case of a thermal bath to perform a reversible driving protocol in the presence of the bath, achieving simultaneously the steps (1\& 2) of the protocol. When resetting the apparatus, reversing the driving protocol allows for getting the invested work back entirely in the limit of quasi-static operation~\cite{latune2024thermodynamically}. In contrast, pure-dephasing remains strictly irreversible and no meaningful reversible quasi-static limit can be obtained in general.

\section{Specific model: a qubit in a cavity}\label{section:qubit-cavity-model}
To push further our analysis, it is useful to examine a specific example of the protocol from Fig.~\ref{fig:measurement-protocol}. We consider that the system is a qubit (e.g. a two-level atom) and the ancilla is a harmonic oscillator (e.g. a cavity). After a suitable qubit-oscillator interaction, measuring excitations in the oscillator reveals the qubit's energy state. This atom-cavity setup allows for the exploration of the impact of measurement strengths, from weak to strong. A key feature of this system, common in the case of weak measurements~\cite{jacobs2006straightforward}, is the large number of possible outcomes, determined by the ancilla's infinite-dimensional space, compared to the qubit observables' two eigenstates.

For simplicity, we first consider that the ancilla is initially in the vacuum (zero temperature) state $\ket{0}$. The qubit and oscillator evolve according to the Hamiltonian:
\begin{equation}
\label{eq:hamiltonian_system-ancilla}
\hat{H}_{S,A}=\underbrace{\frac{\omega}{2}\hat{\sigma}_z\otimes\hat{I}}_{\hat{H}_S}+\underbrace{\hat{I}\otimes\omega_a\hat{a}^{\dagger}\hat{a}}_{\hat{H}_A}+\underbrace{\mu\left(t\right)\hat{\Pi}_g\otimes\left(\hat{a}-\hat{a}^{\dagger}\right)}_{\hat{V}_{SA}},
\end{equation}
where $\omega_a$ is the harmonic oscillator frequency and $\omega$ the qubit frequency. We have introduced the energy eigenbasis $\{\ket{e},\ket{g}\}$ of the qubit, the Pauli matrix $\hat\sigma_z = \ket{e}\bra{e}-\ket{g}\bra{g}$ which corresponds to the measured system observable, and the ancilla's bosonic anihilation (creation) operator $\hat a$ $\left(\hat a^{\dagger}\right)$. Over all this article, we take $\hbar=1$. The coupling operator $\hat V_{SA}$ induces a coherent phase-space displacement of the cavity state if the qubit is in state $\ket{g}$. We assume that the function $\mu\left(t\right)$ has support only in the time interval $(t_0,t_1)$, outside which the coupling is completely switched off.

To explore a broader class of measurements, we allow for an additional direct driving of the ancilla before the pure dephasing occurs. We restrict our analysis to operations resulting in an unconditional phase-space displacement. The resulting unitary evolution then takes the form of a displacement operator
\begin{equation}
\hat{D}\left( \alpha\right)=\exp\left(\alpha\hat{a}^{\dagger}+\alpha^{*}\hat{a}\right).
\end{equation}

Next, the ancilla is put in contact with dephasing environments until it reaches a completely incoherent state in the number basis, at time $t_\text{f}$. The qubit-ancilla state then has the form (see Appendix~\ref{appendix:dynamical_evolution}): 
\begin{eqnarray}
\label{eq:full_system_state_after_dephasing}
\rho_{gg}\left(t_f\right)&=&p_{g0}e^{-|\alpha_1|^2}\sum_{n=0}^{\infty}\frac{|\alpha_1|^{2n}}{n!}\ket{n}\bra{n} \\ \nonumber
\rho_{ge}\left(t_f\right)&=&\rho_{{ge}0}e^{\frac{-\left(|\alpha_1|^2+|\alpha_2|^2\right)}{2}}e^{-i{\phi}}\sum_{n=0}^{\infty}\frac{\alpha_1^n}{\sqrt{n!}}\frac{\alpha_2^{*n}}{\sqrt{n!}}\ket{n}\bra{n}\\ \nonumber
\rho_{eg}\left(t_f\right)&=&\rho_{{eg}0}e^{\frac{-\left(|\alpha_1|^2+|\alpha_2|^2\right)}{2}}e^{i{\phi}}\sum_{n=0}^{\infty}\frac{\alpha_1^{*n}}{\sqrt{n!}}\frac{\alpha_2^n}{\sqrt{n!}}\ket{n}\bra{n}\\ \nonumber
\rho_{ee}\left(t_f\right)&=&p_{e0}e^{-|\alpha_2|^2}\sum_{n=0}^{\infty}\frac{|\alpha_2|^{2n}}{n!}\ket{n}\bra{n},
\end{eqnarray}
where $\phi$ denotes the total phase accumulated during the interaction and driving processes and $\alpha_{1,2}$ the net displacement acquired by the cavity, conditioned on the qubit being in the ground and excited states, respectively. We have denoted $p_{j0}$, $j=e,g$ and $\rho_{ij0}$, $i,j=e,g$ the initial populations and coherences of the qubit. The cavity displacements $\alpha_1$ and $\alpha_2$ associated with states $\ket{g}$ and $\ket{e}$, respectively, are related to the conditional displacement $\alpha_\text{cond}$ generated by the Hamiltonian Eq.~\eqref{eq:hamiltonian_system-ancilla} and the second displacement $\alpha$ via $\alpha_1=\alpha_\text{cond}+\alpha$ and $\alpha_2=\alpha$.
Finally, the ancilla excitation number is read out, and the information is stored in a classical memory.

The conditional state corresponding to the outcome $n$ is
\begin{eqnarray}
\label{eq:conditional_system_state}
p_{g|n}\left(t{_{f}}\right) &=&\frac{p_{g0}}{p_n}e^{-|\alpha_1|^2}\frac{|\alpha_1|^{2n}}{n!} \\ \nonumber
\rho_{ge|n}\left(t{_{f}}\right)&=&\frac{\rho_{{ge}0}}{p_n}e^{\frac{-\left(|\alpha_1|^2+|\alpha_2|^2\right)}{2}}e^{-i{\phi}}\frac{\alpha_1^n}{\sqrt{n!}}\frac{\alpha_2^{*n}}{\sqrt{n!}}\\ \nonumber
\rho_{eg|n}\left(t{_{f}}\right)&=&\frac{\rho_{{eg}0}}{p_n}e^{\frac{-\left(|\alpha_1|^2+|\alpha_2|^2\right)}{2}}e^{i{\phi}}\frac{\alpha_1^{*n}}{\sqrt{n!}}\frac{\alpha_2^n}{\sqrt{n!}}\\ \nonumber
p_{e|n}\left(t{_{f}}\right)&=&\frac{p_{e0}}{p_n}e^{-|\alpha_2|^2}\frac{|\alpha_2|^{2n}}{n!},
\end{eqnarray}
where the conditional state is normalized by the probability of obtaining the outcome $n$, denoted as $p_n$ and defined as
$$p_n=p_{g0}e^{-|\alpha_1|^2}\frac{|\alpha_1|^{2n}}{n!}+p_{e0}e^{-|\alpha_2|^2}\frac{|\alpha_2|^{2n}}{n!}.$$ 

The state update associated with finding $n$ excitations in the ancilla can be formulated in terms of a measurement operator $\hat{M}_n$, according to $\hat{\rho}_{S|n}=\hat{M}_n\hat{\rho}_S\hat{M}_n^{\dagger}/p_n$, where:
\begin{equation}
    \hat{M}_n=\begin{pmatrix}
        e^{-\frac{|\alpha_1|^2}{2}}\frac{\alpha_1^{n}}{\sqrt{n!}}e^{-i\phi} & 0 \\
        0 &  e^{-\frac{|\alpha_2|^2}{2}}\frac{\alpha_2^{n}}{\sqrt{n!}} 
    \end{pmatrix}. 
\end{equation}

When the excitation numbers are grouped into coarse-grained results $r$, the conditional state must be expressed via a map:
\begin{equation}\label{eq:rho_S_given_r}
    \hat{\rho}_{S|r}= \mathcal{E}_r\left[\hat{\rho}_S(0)\right] = \sum_{n\hspace{0.3 mm}\in \hspace{0.3 mm}s_r}\hat{M}_n \hat{\rho}_S(0)\hat{M}^{\dagger}_n,
\end{equation}
where $s_r$ represents the interval of values for $n$ corresponding to the result $r$.

We are particularly interested in the weak measurement limit, attained when the final displacements applied to the ground and excited states are nearly identical. It is therefore convenient to re-express the measurement parameters as:

\begin{eqnarray}
\label{eq:alpha_weak_measurement}
\alpha_1=\overline{\alpha}+\epsilon \\ \nonumber
\alpha_2=\overline{\alpha}-\epsilon.
\end{eqnarray}

The weak measurement regime is expected to occur when $|\epsilon|\ll 1$, such that, only first-order terms in $\epsilon$ are relevant. 

Note that the results of this section correspond to an ancilla initially in the vacuum state, suitable for analytical calculations. Numerical results for thermal state initial ancilla state are provided in Appendix~\ref{appendix:initial_thermal_state}.

\section{Figures of merit for the measurement performance}\label{section:strength_and_efficiency}
\begin{figure} [h!]
    \centering
\includegraphics[width=1.0
\linewidth]{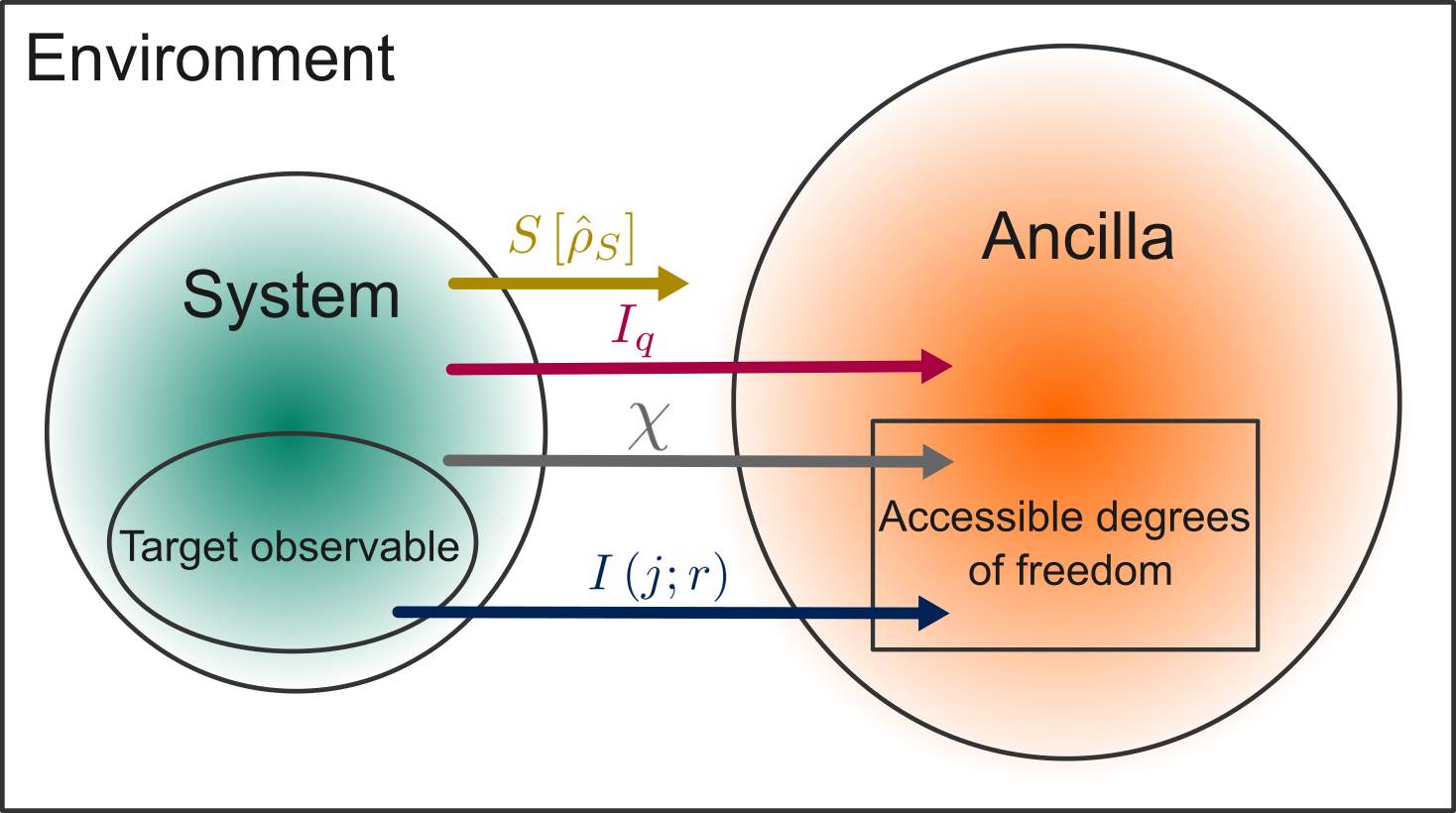}
    \caption{Hierarchy of information flows involved in the figures of merit. The inequality from Eq.~\ref{eq:hierarchy} suggests the following interpretation: From the total amount of information $\log d$ missing to characterize the value of $\hat X$ in the state $\ket{\text{ref}}$, a fraction $S[\hat\rho_S(t_f)]$ is transferred to the ancilla and environment (mustard arrow). A smaller amount $I_q$ is available in the ancilla (pink arrow), and an even smaller amount $\chi$ in the practically accessible degrees of freedom of the ancilla (grey arrow). Finally, only an amount $I({j};{r})$ concerns the target observable (blue arrow).}\label{fig:different_measures}
\end{figure}

In a typical measurement, information is transferred from the system to the ancilla where it can be read. However, this transfer can be incomplete or noisy: a weak (incomplete) measurement leaves information unextracted in the system,  while an inefficient measurement is subject to classical noise, or stores part of the extracted information in inaccessible degrees of freedom, resulting in information loss during the ancilla's readout (see Fig.~\ref{fig:different_measures}). While no standard measures exist to grasp those concepts in the full generality of the zoology of quantum measurements~\cite{wiseman2009quantum}, we build below figures of merit to characterize the performance of the measurements captured by our model.

So as to get quantities characterizing the measurement process, we base our figures of merit on the action of the measurement on a reference pure input system state, chosen to be the one with maximal uncertainty of the measured observable. Namely, we denote in this section the measured observable as $\hat X = \sum_j x_j \ket{j}\bra{j}$, such that the reference state is taken to be $\ket{\text{ref}}= \sum_{j=1}^d \ket{j}/\sqrt{d}$, with $d$ the rank of the measured system observable ($d=2$ in the case of the qubit energy measurement).

We first propose a measure for the strength of the measurement associated with the total information extracted from the system. Starting from our pure reference state $\ket{\text{ref}}$, the latter can be quantified by the von Neumann entropy of the unconditioned output state of the system
$\hat\rho_S(t_f) = \sum_r p_r \hat\rho_{S|r} = \text{Tr}_A \hat\rho_{SA}(t_f)$. ``Unconditioned'' here means averaged over all measurement outcomes, which corresponds to the case where the measurement outcomes are either inaccessible or simply not registered in the classical memory. Indeed, the purity decrease of the average system state is entirely due to correlations built between the system and other degrees of freedom (including the ancilla and its environment) such that it quantifies (or at least upper bounds) the information about the system that can be acquired by measuring those other degrees of freedom. We stress that the von Neumann entropy variation during the measurement depends both on the measurement performance and on the system's initial state (e.g. if it is already in a maximally mixed state, the entropy does not increase further); hence, the necessity of a reference state. In summary, we quantify the strength of the measurement via:\begin{eqnarray}
\label{eq:strength_measure}
\xi&=&\frac{S\left[\hat{\rho}_S(t_f)\right]}{\log{d}},
\end{eqnarray}
where it is understood that $\hat\rho_S(t) = \sum_r p_r {\cal E}_r[\ket{\text{ref}}\bra{\text{ref}}]$, and $S\left(\hat{\rho}\right)$ denotes the von Neumann entropy of the density operator $\hat{\rho}$. We have normalized the strength measure $\xi$ so that $\xi=0$ means an absence of extracted information, while $\xi=1$ indicates maximal extraction of the observable's statistics, leaving the system in a completely mixed state of the eigenstates $\ket{j}$. When $\xi \ll 1$, the measurement is weak; at $\xi=1$, it is strong, with intermediate values providing a scale to rank general measurements.

The measurement performance is also limited by the information ``lost'' in the environment (that is, which cannot be retrieved in the accessible degrees of freedom of the ancilla, but could be read in principle in other degrees of freedom of the ancilla or in the environment). 
Still considering that the system starts in the reference state $\ket{\text{ref}}$, this loss can be quantified from the purity (or von Neumann entropy) of the conditional system state $\rho_{S|r} = \text{Tr}\{\rho_{SA|r}\}$ associated with obtaining the result $r$. Indeed, a pure conditional state means that reading the ancilla allowed the observer to obtain all the information transferred during the system-ancilla interaction, while a mixed conditional state means that more information could be obtained by reading the inaccessible degrees of freedom.
To obtain a quantity independent of both the measurement outcome and the measurement strength, and maximal when all available information is accessed, we use the Holevo information \cite{Wilde2011Jun} $\chi = S[\hat\rho_S]-\sum_r p_r S(\hat\rho_{S|r}) \geq 0$.  which takes its maximum value $S[\hat\rho_S]$ when all the extracted information about $S$ can be read in the ancilla. Normalizing to this maximum value leads to the following figure of merit for the measurement efficiency:
\begin{equation}
\label{eq:efficiency_purity_conditional_state}
    \eta=\frac{S\left(\hat{\rho}_S\right)-\sum_r p_r S\left(\hat{\rho}_{S|r}\right)}{S\left(\hat{\rho}_S\right)} = \frac{\chi}{S\left(\hat{\rho}_S\right)}.
\end{equation}
This quantity generalizes the notion of quantum efficiency for detectors in weak continuous measurement scenarios~\cite{jacobs2006straightforward,Wiseman1993Jan, ficheux2018dynamics}, and can be identified from the stochastic master equations governing the dynamics under continuous measurement.

The efficiency $\eta$ is equal to $0$ (resp. $1$) when none of the information (resp. all the information) extracted about the system is obtained by reading the ancilla. In the absence of coarse-graining (that is, $p(r|n)=\delta_{r,n}$), $\chi$ is equal to the quantum mutual information $I_q$ between the ancilla and the qubit at time $t_f$. In the presence of coarse-graining, $\chi\leq I_q$. However, in any case, $\chi$ is an upper bound on the correlations between the measurement outcome distribution $p(r)$ and the statistics of any observable of $S$~\cite{Wilde2011Jun} (see also Appendix.~\ref{appendix:hierarchy_information}).

While the efficiency measure does quantify the ability of our measurement to actually make information about the system available in the ancilla, it is not always sufficient to eliminate bad quality measurements.

Indeed, as illustrated in the next section, there are cases where the efficiency and the strength are maximal $(\xi=1,\eta=1)$, meaning that maximal correlations have built between the system and the accessible degrees of freedom of the ancilla, and yet the measurement result actually does not bring any information about the target measurement observable, or even about any system observable at all. This effects occurs whenever the ancilla is not read in the optimal basis, as noted e.g. in superconducting qubit weak measurement experiments \cite{murch2013observing}.

While this issue is avoided in well-calibrated measurement setups, it illustrates the necessity an additional figure of merit in order to fully characterize the quality of arbitrary measurements. We use the classical mutual information $I\left(\{j\};\{r\}\right) = \sum_{j,r}p_{j,r}\log(p_{j,r}/p_jp_r)$ between the classical output of the measurement, the result $r$, and the target observable statistics, initially by distribution $p_j = \bra{j}\hat\rho_S(t_f)\ket{j}$. It is computed from the joint distribution $p_{j,r}=\bra{j}\hat\rho_{S|r}\ket{j}$ and $p_r=\sum_n p_n p(r|n)=\text{Tr}\{{\cal E}_r[\hat\rho_S(0)]\}$. The Holevo bound~\cite{Preskill16,Wilde2011Jun} ensures that $I\left(\{j\};\{r\}\right)\leq \chi$ (see Appendix.~\ref{appendix:hierarchy_information}, such that a natural normalization for our third figure of merit $\eta_{X:r}$ is:

\begin{equation}
\label{eq:efficiency_mutual_information}
    \eta_{X:r}=\frac{I\left(\{j\};\{r\}\right)}{S\left(\hat{\rho}_S\right)-\sum_r p_r S\left(\hat{\rho}_{S|r}\right)}.
\end{equation}

This quantity vanishes when the measurement result $r$ and the observable value $x_j$ are two independent random variables; it reaches $1$ when all the system-ancilla correlations are accounted for by the classical correlations in $p_{j,r}$. 

It is also worth mentioning that, unlike the efficiency measure $\eta$, the quantity $\eta_{X:r}$ is not independent of the measurement strength, and this dependence cannot be simply removed by renormalization. $\eta_{X:r}$ is therefore not equivalent to $\eta$, and rather brings additional information about the measurement efficiency.
Together, these three figures of merit allow us to analyze the performance of nonideal measurements of a system observable, as we do in the following for our qubit-cavity model. 

We finish by two remarks. First, owing to the chosen normalizations, the product of the three figures of merit verifies 
\be\label{eq:productmeasures}
\xi \eta \eta_{X:r}=\frac{I\left(\{j\};\{r\}\right)}{\log{d}},
\ee
 which corresponds to the ratio between the information available in the measurement result $r$ about the target observable, over the total initial uncertainty (the Shannon entropy of the initial distribution $p_j(0)=1/\sqrt{d}$). This product goes to $1$ for a projective (strong and efficient) measurement. Second, the information quantities involved in the figures of merit fulfill the hierarchy 
\be\label{eq:hierarchy}
\log d \geq S[\hat\rho_S(t_f)]\geq I_q\geq \chi\geq I(\{j\},\{r\})
\ee
which supports our interpretation in terms of information loss for the different sources of nonideality in the measurement process (see Fig.~\ref{fig:different_measures}).

\section{Analysis of a single measurement: efficiency, strength and work}\label{section:single_measurement}
In this section, we use the three figures of merit introduced above to analyze the measurement generated by the model of section~\ref{section:qubit-cavity-model}. This model contains two key parameters: $\bar{\alpha}$, the average cavity displacement, and $\epsilon$, the displacement conditional to the qubit's energy.

To evaluate the strength measure $\xi$, we compute the average reduced system state:
\be \label{eq:rho_Sexample}
\hat\rho_S(t_f)=\begin{pmatrix}
    p_{e0} & \rho_{eg0}e^{i\phi-2\epsilon^2}\\
    \rho_{eg0}^*e^{-i\phi-2\epsilon^2} & p_{g0}.
\end{pmatrix}
\ee

\begin{figure} [h!]
    \centering
\includegraphics[width=0.7
\linewidth]{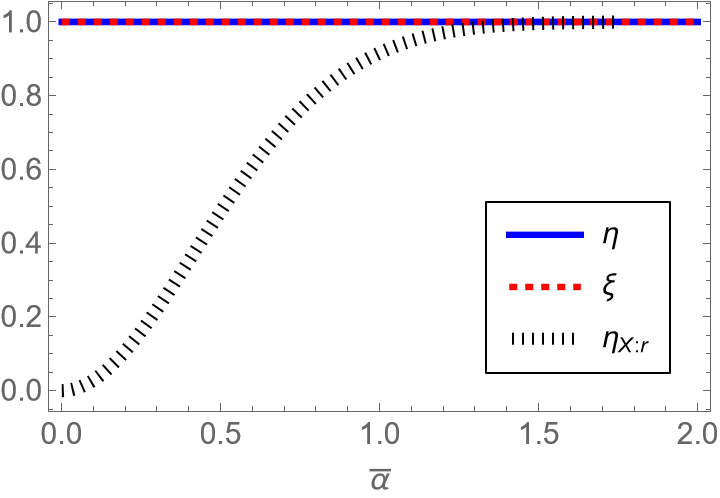}
    \caption{Strength ($\xi$), efficiency ($\eta$), and normalized mutual information  ($\eta_{X:r}$) are examined in terms of $\bar{\alpha}$. While both efficiency and strength remain constant at 1, no information is extracted from the measurement protocol. Therefore, normalized mutual information ($\eta_{X:r}$) should be utilized to quantify the measurement quality. The initial qubit state is $\frac{1}{\sqrt{2}}\left(\ket{g}+\ket{e}\right)$.}\label{fig:efficiency_and_strength}
\end{figure}

From Eqs.~\eqref{eq:strength_measure} and~\eqref{eq:rho_Sexample}, we see that $\xi$, plotted in Fig.~\ref{fig:efficiency_and_strength}, only depends on $\epsilon$. Moreover, $\rho_S(t_f)$ is unaffected by the coarse-graining of the result, and so is $\xi$. Notably, when $\epsilon$ reaches around 1.5, the measurement becomes strong, generating maximal correlations between the qubit and the other parts of the full system, and leaving the qubit in a completely mixed average state.

\begin{figure}[htp]
    \begin{center}
    \includegraphics[width=6cm]{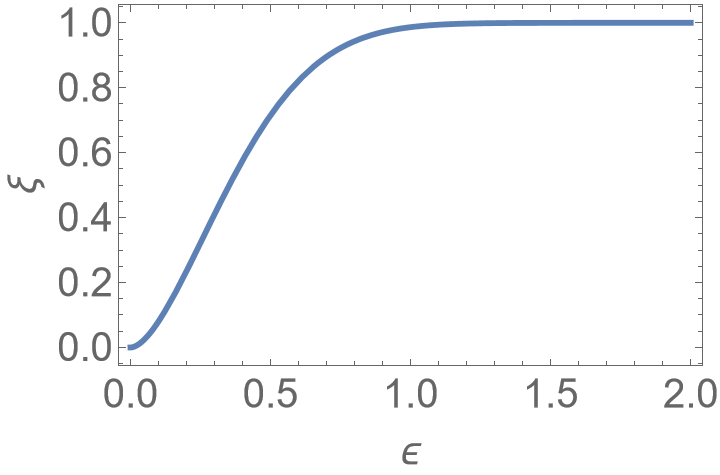}
    \end{center}
    \caption{Strength, $\xi$, as a function of $\epsilon$ for the qubit measurement model. The strength depends only on $\epsilon$ and converges to 1. The qubit was initialized in the reference state $\ket{\text{ref}}=\frac{1}{\sqrt{2}}\left(\ket{g}+\ket{e}\right)$.}
\label{fig:strength_in_terms_of_epsilon}
\end{figure}

\begin{figure}[htp]
    \centering
    \includegraphics[width=6cm]{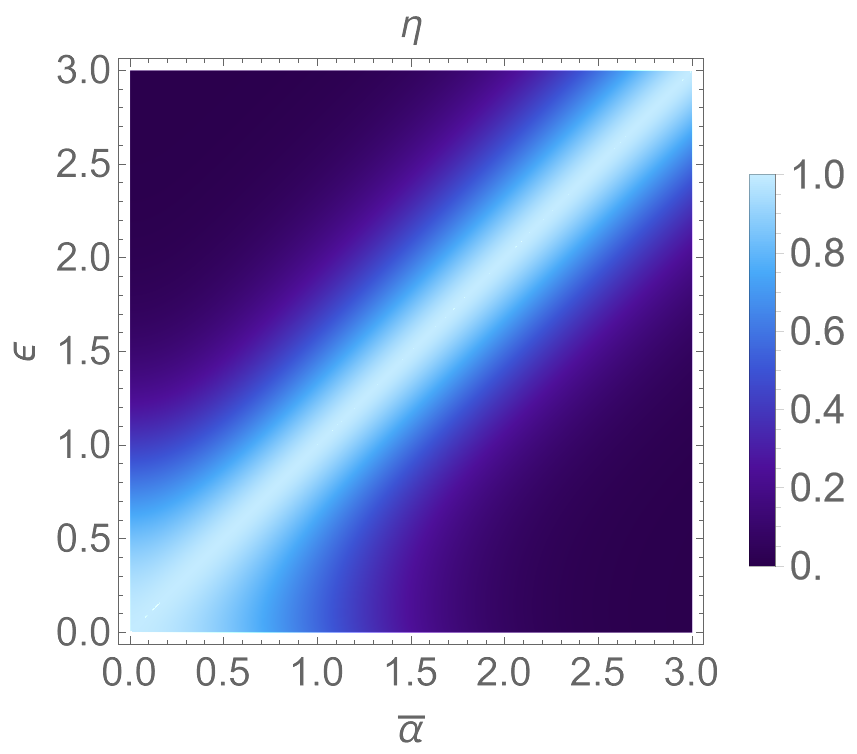}
    \caption{Measure of efficiency $\eta$ in terms of $\bar{\alpha}$ and $\epsilon$ in the coarse-grained qubit measurement scenario where the two measurement results correspond to the total absence of excitation in the field and the presence of one or more excitations, respectively. The qubit was initialized in the reference state  $\ket{\text{ref}}=\frac{1}{\sqrt{2}}\left(\ket{g}+\ket{e}\right)$.}
\label{fig:eta_coarse_graining}
\end{figure}

In the absence of coarse-graining, and for a pure ancilla initial state, the measure of efficiency $\eta$, defined in Eq.~\eqref{eq:efficiency_purity_conditional_state}, is equal to 1~\footnote{This remains true for any model as those conditions imply that the conditional post-measurement state is captured by a single measurement operator $\hat\rho_{S|r}\propto \hat M_r\hat\rho_S(0)\hat M_r^\dagger$}. We analyze here the impact of coarse-graining by analyzing the case where the measurement result takes two values, corresponding to no excitation and any number of excitations in the ancilla, respectively (e.g. as a photodiode would do). This scheme is practically motivated for weak measurements, as a weak (short) qubit-cavity interaction makes high numbers of excitations very unlikely. This is implemented by two projectors $\hat\Pi_{r=e}=\ket{0}\bra{0}$ and $\hat\Pi_{r=g}=\sum_{n>0} \ket{n}\bra{n}$. In this case, the efficiency measure $\eta$ depends on $\bar{\alpha}$ and $\epsilon$. As shown in Fig.~\ref{fig:eta_coarse_graining}, efficiency peaks when $\epsilon$ and $\bar{\alpha}$ are close to each other, as the conditional field displacement associated with the qubit being in the excited state becomes negligible with respect to the one associated with the ground state (and finding excitations in the cavity is strongly correlated to the qubit being in the ground state). Specifically, when $\bar{\alpha} = \epsilon$, the efficiency reaches 1. At this point, the measurement is extremely asymmetric as outcome $r=1$ is associated with the qubit excited state with certainty, while $r=0$ only provides partial information about the qubit state. As a consequence, $p(r=e) = p_e+p_g e^{-4|\bar \alpha|^2}\neq p_e(0)$, meaning that the measurement is {\it biased} unless $\vert\bar\alpha\vert \gtrsim 1$~\cite{guryanova2020ideal}.

\begin{figure}[htp]
    \centering
    \includegraphics[width=6cm]{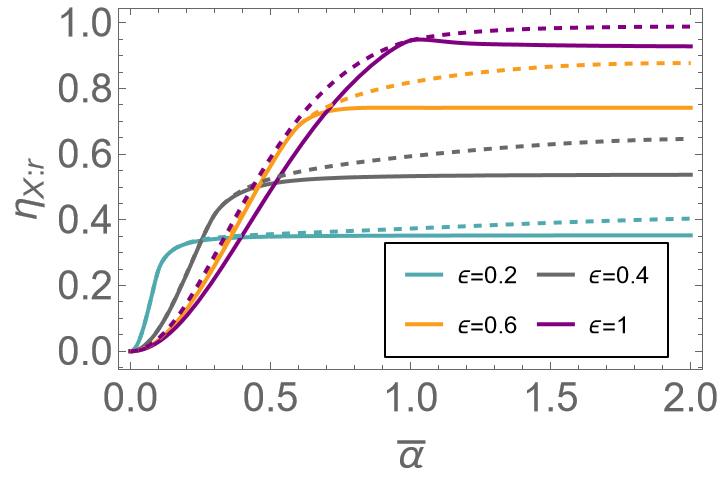}
    \caption{Figure of merit based on the classical mutual information $\eta_{X:r}$, for the measurement observable $\hat X=\hat\sigma_z$, as a function of $\bar{\alpha}$ and at different values of $\epsilon$, for the qubit measurement model. Dashed lines (resp. solid line) correspond to the case with coarse-grained (without coarse-graining). The qubit was initialized in the reference state  $\ket{\text{ref}}=\frac{1}{\sqrt{2}}\left(\ket{g}+\ket{e}\right)$.}
\label{fig:epetaI}
\end{figure}

Fig.~\ref{fig:epetaI} presents the figure of merit  $\eta_{X:r}$ based on classical mutual information as a function of $\bar{\alpha}$ for various values of $\epsilon$. We compare cases without coarse-graining (solid lines) and with coarse-graining (dashed lines). Without coarse-graining, the mutual information for each $\epsilon$ reaches a plateau, whose value grows with $\epsilon$, highlighting its dependence on measurement strength. This plateau emerges around $\bar{\alpha} = \epsilon$. With coarse-graining, $\eta_{X:r}$ increases until it reaches a higher plateau than in the fine-grained case.

\begin{figure}[htp]
    \centering
    \includegraphics[width=6cm]{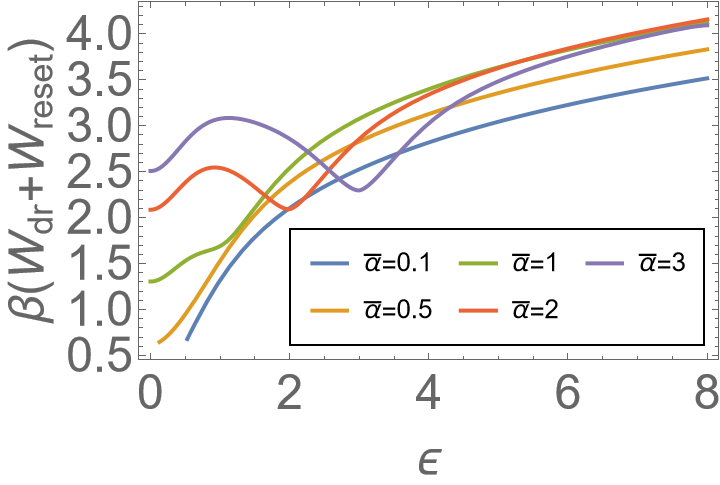}
    \caption{Lower bound on the work cost of the qubit measurement as a function of $\epsilon$ for different values of $\bar{\alpha}$. The qubit was initialized in the reference state $\ket{\text{ref}}=\frac{1}{\sqrt{2}}\left(\ket{g}+\ket{e}\right)$.}
\label{fig:work_in_terms_of_epsilon_for_different_alpha}
\end{figure}

We now discuss the work cost associated with the measurement. As discussed in section~\ref{section:energetics}, the lower bound on the work is unaffected by coarse-graining as long as the coarse-grained subspaces are orthogonal, as assumed here. The minimum work is plotted in Fig.~\ref{fig:work_in_terms_of_epsilon_for_different_alpha} as a function of $\epsilon$ for different values of $\bar{\alpha}$. For low $\bar{\alpha}$, the work bound increases monotonically with $\epsilon$. When $\bar{\alpha} > 1$, the behavior becomes non-monotonous. A minimum appears at $\epsilon = \bar{\alpha}$, where the bound can drop below the value for $\bar{\alpha}\to 0$. This behavior arises because of the peculiar form of the conditional qubit-ancilla state at $\bar{\alpha} = \epsilon$:
\be
p_{g|n}(t_f)&=& \frac{p_{g0}}{p_n}e^{-4\epsilon^2}\frac{\epsilon^{2n}}{n!}\\
\rho_{ge|n}(t_f)&=&\frac{\rho_{ge0}}{p_n}e^{-i\phi-2\epsilon^2}\frac{(2\epsilon)^{2n}}{\sqrt{n!}}\delta_{n,0}\nonumber\\
\rho_{eg|n}(t_f)&=&\frac{\rho_{eg0}}{p_n}e^{i\phi-2\epsilon^2}\frac{(2\epsilon)^{2n}}{\sqrt{n!}}\delta_{n,0}\nonumber\\
p_{e|n}(t_f)&=&\frac{p_{g0}}{p_n}\delta_{n,0},\nonumber
\ee
where the excitation number distribution associated with the excited state is a Kronecker delta and has zero entropy. For large $\bar{\alpha}$ such that the excitation number distribution associated with the ground state is peaked around $n~|\bar\alpha|^2\gg 1$, the Shannon entropy of the total excitation number distribution scales like the sum of the Shannon entropies of the two conditional distributions $p(n|e)$ and $p(n|g)$, and therefore drops when $\bar\alpha\simeq \epsilon$ (see also Appendix~\ref{appendix:general_strong_versus_weak_measurement_cost} for a more general discussion of this behavior). This effect is not visible for $\bar\alpha \ll 1$ as $p(0|g)$ becomes non-negligible. In this protocol, the work bound clearly depends on the measurement strength. We note, however, that when using an ancilla with a bounded Hilbert space, such as a qubit, the work cost for a single measurement can be independent of the strength for specific initial system states~\cite{mancino2018entropic}.

We finish by mentioning the effect of a non-zero temperature of the ancilla. As shown in Appendix~\ref{appendix:initial_thermal_state}, initializing the ancilla in a thermal state leads to reduced values of the measurement efficiency $\eta$ of the product $\xi\eta\eta_{X:r}$ (in contrast, $\xi$ and $\eta_{X:r}$ may increase), for given values of $\bar\alpha$ and $\epsilon$. The measurement extracts less useful information, but the lower bound on the work cost decreases.

\begin{figure} [h!]
    \centering
\includegraphics[width=0.8
\linewidth]{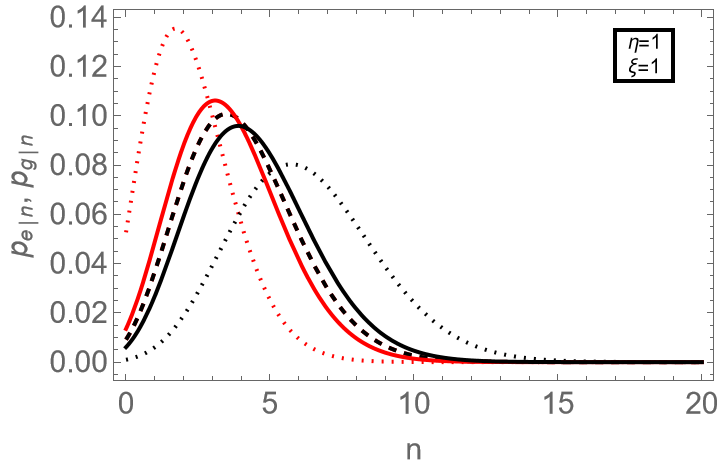}
    \caption{$p_{e|n}$ (red) and $p_{g|n}$ (black) versus the number of field excitations $n$ for three values of $\bar{\alpha}$ with $\epsilon=2$. The dashed line corresponds to $\bar{\alpha}=0$, the solid line represents $\bar{\alpha}=0.1$ and the dotted line to $\bar{\alpha}=0.5$. For $\bar{\alpha}=0$, both lines overlap. The qubit was initialized in the reference state $\ket{\text{ref}}=\frac{1}{\sqrt{2}}\left(\ket{g}+\ket{e}\right)$.}
    \label{fig:pe_and_pg}
\end{figure}

Finally, we illustrate in this model that, as explained in the previous section, the measures of strength (related entropy increase of the average state) and efficiency (related to the entropy increase of the conditional state) are not always sufficient to fully assess the measurement performance. Consider the case without coarse-graining such that the measurement result is one of the excitation numbers $r=n \in \mathbb{N}$. We plot in Fig.~\ref{fig:pe_and_pg} the probability distributions for the excited and ground states (corresponding to two possible outcomes), for $\epsilon=2$ (ensuring $\xi_S=\eta=1$) and different values of $\bar\alpha$. When $\bar\alpha=0$, both distributions perfectly overlap (dashed black line). This indicates that although complete information about the system state is transferred to the ancilla (the conditional system state $\rho_{S|n}$ is pure whatever $n$), the outcome does not necessarily encode any meaningful information about the target measurement observable $\sigma_z$. Actually, in this case, the conditional qubit state verifies $p_{g|n}=p_{g|0}$, $p_{e|n}=p_{e|0}$, $\rho_{eg|n} = (-1)^n\rho_{eg|0}$. The backaction associated with finding $r=n$ is only a phase shift, preserving purity of the conditional qubit state, but decreasing that of the average state. In this case, the measurement does not bring information on a qubit observable, but instead detects phase jumps.

As shown in Fig.~\ref{fig:efficiency_and_strength}, the normalized classical mutual information $\eta_{\sigma_z:r}$ increases with $\bar{\alpha}$, indicating that more information about $\sigma_z$ is being extracted. In contrast, both the detection efficiency and strength measures remain fixed at 1 and, in this case, are independent of $\bar{\alpha}$.

\begin{figure} [h!]
    \centering
\includegraphics[width=0.8
\linewidth]{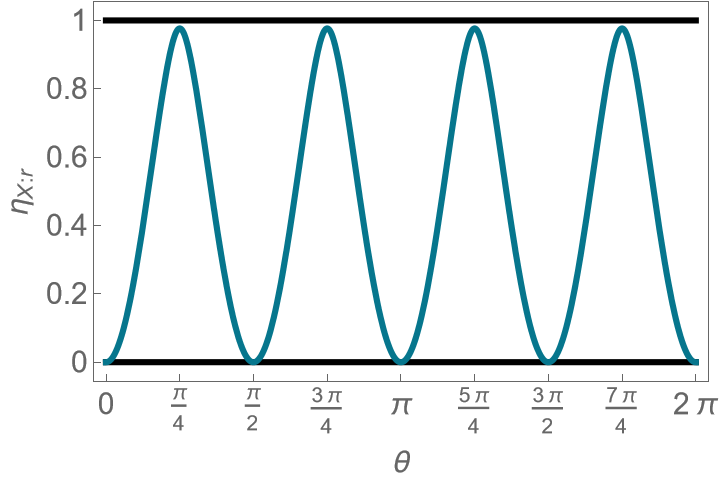}
    \caption{Normalized mutual information $\eta_{X:r}$ as a function of the angle $\theta$ of the unitary evolution described by Eq.~\eqref{eq:unitaryrot}, for $\bar{\alpha}=0$ (black) and $\epsilon=5$ (turquoise). The qubit was initialized in the reference state  $\ket{\text{ref}}=\frac{1}{\sqrt{2}}\left(\ket{g}+\ket{e}\right)$.}\label{fig:Imut_unitary}
\end{figure}

One might wonder if the measurement protocol can be corrected to bring information about the target observable when $\bar\alpha=0$. Noting that the information about $\sigma_z$ is encoded in the phase of the ancilla at the end of step 1, we consider applying on the ancilla, before the pure dephasing takes place, the unitary evolution described by, for all $n$:
\begin{eqnarray}\label{eq:unitaryrot}
    \ket{2n}&\rightarrow&\cos{\theta}\ket{2n}+\sin{\theta}\ket{2n+1} \\ \nonumber
    \ket{2n+1}&\rightarrow&\cos{\theta}\ket{2n+1}-\sin{\theta}\ket{2n}. 
\end{eqnarray}

As shown in Fig.~\ref{fig:Imut_unitary}, this operation indeed increases the classical mutual information between the ancilla excitation number $n$ and the value of $\hat\sigma_z$. Maximal values $\eta_{\sigma_z:r}=0.977$ are obtained for $\theta= \frac{\pi}{4}+m\frac{\pi}{2}$, where $m$ is a natural number. Further improvement might be obtained owing to a global unitary on the ancilla.

\section{Concatenated weak measurement versus strong measurement}\label{section:measurement_sequence}
In this section, we compare the work cost associated with a strong measurement obtained by concatenating a long sequence of weak measurements, versus a single iteration of the measurement procedure of section~\ref{section:qubit-cavity-model} with large strength. When the duration of each weak measurement involved in the sequence goes to zero, a so-called weak continuous measurement is obtained~\cite{jacobs2006straightforward}. We assume that, at each iteration, the whole process in Fig.~\ref{fig:measurement-protocol} is performed, including the reset of the ancilla and classical memory. The system state is incremented according to Eq.~\eqref{eq:rho_S_given_r}, with $\hat{M}_n$ being infinitesimally close to $\idop$. This cycle continues until all relevant information is extracted from the system. 

\begin{figure}[htp]
    \centering
    \includegraphics[width=6cm]{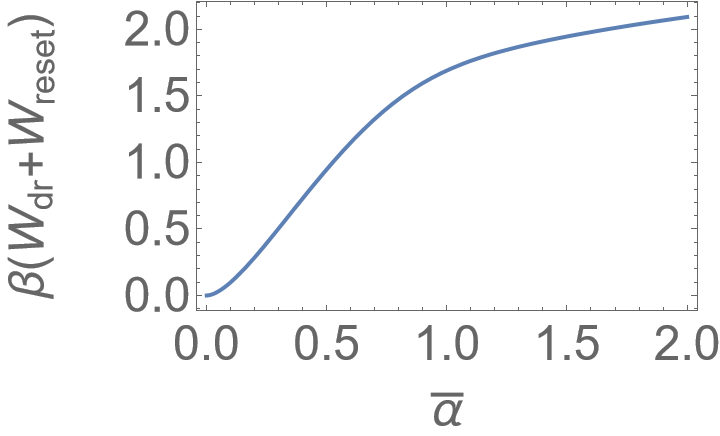}
    \caption{Work bound as a function of $\bar{\alpha} = \epsilon$, for a single qubit measurement with $p_{e0} = p_{g0}= \rho_{ge0}=\rho_{eg0}=\frac{1}{2}$. The work bound has a lower slope at small and large $\bar{\alpha}$.}
\label{fig:work_in_terms_of_alpha}
\end{figure}

From the analysis of a single measurement in the previous section, the minimum work cost occurs at $\bar{\alpha} = \epsilon$ (see Fig.~\ref{fig:work_in_terms_of_epsilon_for_different_alpha}). We therefore assume this equality to hold hereafter, and analyze results in terms of $\bar\alpha$. We plot the work bound as a function of $\bar{\alpha}=\epsilon$ in Fig.~\ref{fig:work_in_terms_of_alpha}, showing that it increases sub-linearly with $\bar{\alpha}$. This behavior provides a first indication that $N \gg 1$ weak measurements of strength $\xi_\text{weak}\ll 1$ should have a larger work cost than a single measurement of maximal strength $\xi_\text{strong} \sim N\xi_\text{weak} \sim \log 2$. We formalize this result in the remainder of this section, showing that a sequence of weak measurement costs more work than a direct implementation of the resulting strong measurement, and that this statement also applies to the other measurement figures of merit.

To this goal, we analyze the behavior of a sequence of weak measurements as a single stronger measurement. We thus need a rule to concatenate the $N$ measurement outcomes $\vec r=(r_1,....,r_N)$ obtained in the elementary measurements of the sequence into a single result $R=e,g$, and derive the compute qubit state update. Due to the asymmetry (and resulting bias) of the measurement associated with our example (see section~\ref{section:single_measurement}), collapsing towards state $\ket{e}$ and $\ket{g}$ typically requires a different number of weak measurements, with relative weights depending on the parameters $\bar\alpha$ and $\epsilon$. As a consequence, combining the weak measurement results of a sequence $\vec r$ into a single result $R$ is not trivial. We therefore adopt a Bayesian systematic approach. Namely, we assign to $R$ the eigenstate (ground or excited) with highest probability, given the sequence $(r_1,....,r_N)$, as computed from Eq.~\eqref{eq:rho_S_given_r} via
\be\label{eq:pcond_Bayesian}
p(j|r_N,....,r_1) = \frac{\bra{j}{\cal E}_{r_N}\otimes...\otimes{\cal E}_{r_1}[\rho_S(0)]\ket{j}}{\text{Tr}\{{\cal E}_{r_N}\otimes...\otimes{\cal E}_{r_1}[\rho_S(0)]\}},\quad j=e,g.
\ee

\begin{figure}[htp]
    \centering
    \includegraphics[width=9cm]{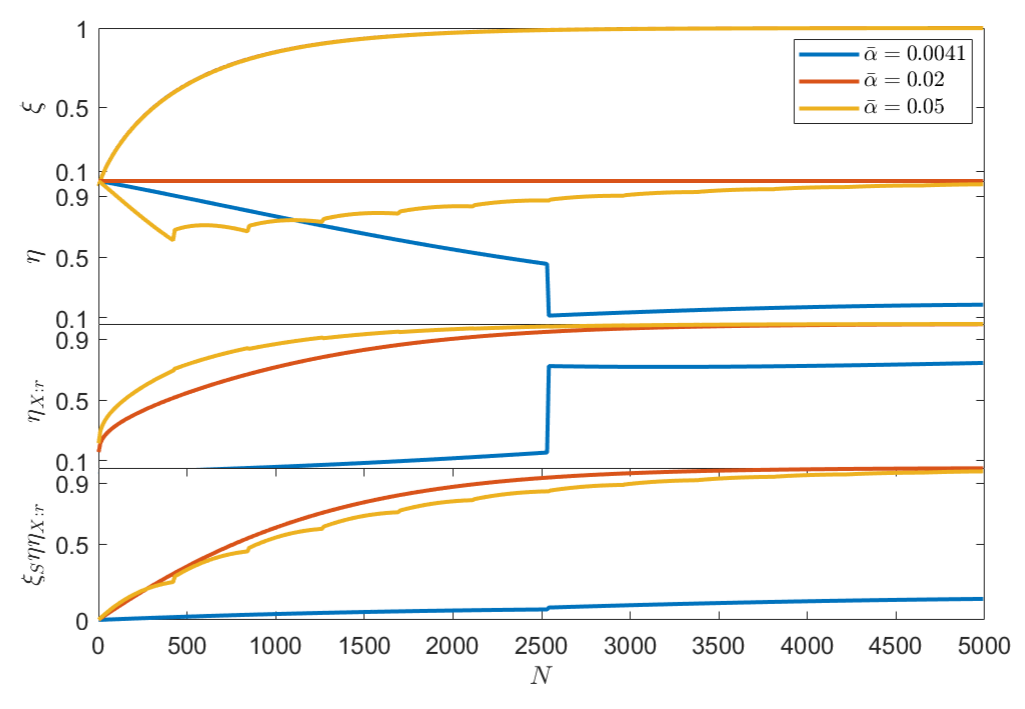}
    \caption{Strength ($\xi$), detection efficiency ($\eta$), normalized classical mutual information ($\eta_{X:r}$), and product $\xi\eta\eta_{X:r}$ as a function of the number of iterations ($N$) for different values of average displacement ($\bar{\alpha}$). The sudden jump in detection efficiency and normalized classical mutual information arises due to the concatenation rule. This occurs when an additional outcome combination sequence satisfies $p_{e|\Vec{r}}> p_{g|\Vec{r}}$. The qubit was initialized in the reference state  $\ket{\text{ref}}=\frac{1}{\sqrt{2}}\left(\ket{g}+\ket{e}\right)$.}
\label{fig:panel_after_several_repetitions}
\end{figure}

We now analyze the figures of merit for the measurement emerging from such concatenation procedure. They are plotted in Fig.~\ref{fig:panel_after_several_repetitions} as a function of the number $N$ of iterations in the sequence.

The strength measure (top panel) is independent of $\bar{\alpha}$. For $\epsilon=0.02$, the value used for Fig.~\ref{fig:panel_after_several_repetitions}, the strength reaches 1 after approximately {3000} iterations.

The detection efficiency measure $\eta$ is plotted in the second panel. While a single measurement has maximal detection efficiency ($\eta=1$) in the absence of coarse-graining, repeated protocol applications reduce the purity of the conditional ancilla states, making efficiency dependent on $\bar{\alpha}$ and $\epsilon$. This can be intuitively understood noting that the Bayesian procedure to combine the sequence results into a two-valued outcome is a kind of coarse-graining.  For sufficient large values of $\bar\alpha \gtrsim \epsilon$, the efficiency decreases to minimum values over a few hundreds of repetitions, before converging to $1$ at large number of repetitions (it remains all along at exactly $1$ for $\bar\alpha=\epsilon$). For $\bar\alpha\ll\epsilon$, the efficiency drops and convergence is not ensured.

The third panel in Fig.~\ref{fig:panel_after_several_repetitions} shows $\eta_{X:r}$ as a function of the number of measurements in the sequence. Convergence to $1$ at large numbers is obtained for sufficiently large $\bar\alpha \gtrsim\epsilon$. Finally, the bottom panel shows the product $\xi_S\eta\eta_{X:r}$ (see Eq.~\eqref{eq:productmeasures} and related discussion), demonstrating convergence towards a projective measurement for long enough measurement sequences, and $\bar\alpha \gtrsim\epsilon$.

\begin{figure}[htp]
    \centering
    \includegraphics[width=8cm]{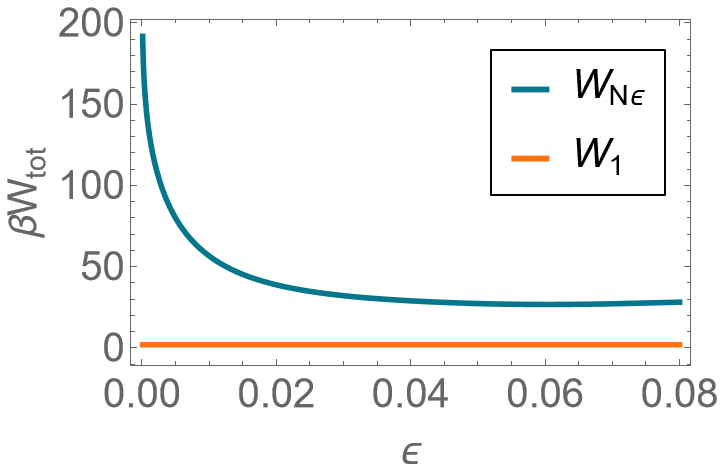}
    \caption{Total work bound to obtain a strong efficient measurement by concatenating weak measurements (turquoise line) of value $\epsilon$ versus a strong measurement (orange line) for $\epsilon=1.5$  with a detection efficiency and normalized mutual information of at least 0.999. The used data is $p_{e0}=p_{g0}=\rho_{eg0}=\rho_{ge0}=\frac{1}{2}$ and $\bar{\alpha}=\epsilon$.}
\label{fig:pWproj_2}
\end{figure}

We end this section by comparing the work bound $W_1$ associated with a single strong measurement, characterized by $\epsilon_\text{strong}=\bar\alpha_\text{strong} = 1.5$ (achieving all three figures or merit to be larger or equal to 
$0.999$), versus the total work $W_{N\epsilon}$ associated to a sequence of $N$ weak measurements characterized by $\epsilon=\bar\alpha \ll 1$. The value of $N$ in the comparison is chosen such that the resulting measurement achieves the same performance, that is all three figures of merits larger or equal to $0.999$. As shown in Fig.~\ref{fig:pWproj_2} the work cost of the measurement sequence (blue line) is typically hundred times higher than the single strong measurement (orange line), and this remains true for a large range of values of $\epsilon=\bar\alpha$.

As we argue below, this trend is actually very general. As detailed in Appendix~\ref{appendix:general_strong_versus_weak_measurement_cost}, for any protocol where dephasing plays the role of converting quantum states into classical ones -- a mechanism that aligns with the framework of quantum Darwinism~\cite{zurek2009quantum} --, and largely regardless of the specific model, initial ancilla state, and many other details, we show that the work cost for $N$ concatenated weak measurements is significantly higher than that of a single strong, efficient measurement achieving similar performance. Our proof is based on phenomenological assumptions about the functional form of the ancilla's state distribution $p_n$ that must be satisfied for a strong and weak measurement, respectively.

We find that the lower bound of the work cost for a sequence of weak measurements generating a strong measurement under concatenation must scale as $W_\text{weak}\propto 1/\epsilon_\text{weak}^2$, for $\epsilon_\text{weak}\ll 1$. In contrast, for a single strong measurement, the lower bound typically scales as $W_\text{strong} \propto K(l\log(2\pi\epsilon_\text{strong})+1/2))$, for $\epsilon_\text{strong} = {\cal O}(1)$ and $K$ the number of possible outcomes of the strong measurement (i.e., the number of eigenstates of the measured system observable), and $l \in \{0,1\}$ a model-dependent coefficient.
To examine weak continuous measurements, we introduce $a = \epsilon_\text{strong}/\epsilon_\text{weak}$, and we consider the limit $a \gg 1$, $\epsilon_\text{strong}={\cal O}(1)$. It is straightforward to see that $W_\text{weak}\propto a^2/\epsilon_\text{strong}^2$ becomes much larger than $W_\text{strong}$ (independent on $a$), in this limit.
We stress that this result is independent on the chosen efficiency figures of merit and constitutes the central result of this article. 

Finally, we note that those results are in agreement with the experimental observations of \cite{mancino2018entropic}, where the cost of a single qubit measurement (more precisely, its entropy production) is shown to be independent of the measurement strength for their protocol, when starting in the same reference state $\ket{\text{ref}}$. As a consequence, the work cost for a sequence of $N ~ 1/\epsilon_\text{weak}^2 \gg 1$ weak measurements is expected to be $N$ times larger.

\section{Conclusions}
This article examines a minimal setup for modeling non-ideal quantum measurements, consisting of an interacting system of interest and an ancilla. After the interaction, the ancilla undergoes dephasing through its interaction with other degrees of freedom of the apparatus, transforming quantum data into classical outputs in the form of a SBS state, i.e. a classically correlated state of the ancilla and the system, simulating the objectification process. The ancilla is read out, and the data is stored in a classical memory. Finally, both the ancilla and the classical memory are reset to restart the measurement protocol.

First, we analyzed the energy balance of the general model with dephasing. We concluded that the work bound is higher when dephasing occurs than when replaced by a continuously interacting dissipative bath during the measurement protocol.

Next, we focused on a specific case of this model where the ancilla is a quantum harmonic oscillator and the system is a qubit. We analyzed the performance of this protocol using three metrics: strength, detection efficiency, and normalized classical mutual information. While detection efficiency and strength might seem sufficient to quantify measurement performance, we identified instances where both were maximal yet no information was extracted about the observable of interest. In such cases, the normalized classical mutual information becomes essential, despite its dependence on measurement strength. These metrics represent the information extracted from the system (strength), the information reaching the accessible degrees of freedom of the ancilla (detection efficiency), and the information about the observable of interest that arrives at the accessible degrees of freedom of the ancilla (normalized classical mutual information). The product of these three measures indicates the total information gained from the measurement protocol.

Finally, we studied the energy balance of this protocol in detail. We found that the work bound required for a strong, efficient measurement is much lower when achieved with a single measurement compared to a series of concatenated weak measurements. 

This article demonstrates that the minimal measurement setup can model measurements from weak to strong and from completely inefficient to fully efficient, all without requiring any projective measurement. Additionally, we conclude that a series of weak measurements is generally less cost-effective than a single efficient strong measurement. We hope this conclusion will assist experimentalists in selecting the optimal measurement protocol and improving the work cost of these protocols.
\section*{Acknowledgments}
This project was supported by the French National Research Agency (ANR) under grant ANR-22-CPJ1-0029-01. The authors extend their gratitude to Yashovardhan Jha, Camille L Latune and Paul Skrzypczyk for fruitful discussions that contributed to this work.  
\clearpage
\onecolumngrid
\appendix
\section{Dissipation energetic calculations}\label{appendix:work_bound_dissipation}
In this appendix, we calculate the work bound for the scenario where dephasing in our model is replaced by dissipation. Unlike the dephasing model, driving and action of the bath can happen simultaneously \cite{latune2024thermodynamically}. We will analyze the work cost of the protocol step by step. 

Similar to the dephasing case, we assume that at the start of the protocol ($t_0$), the ancilla is in a free state in equilibrium with its environment.

\underline{Drive in presence of thermalizing bath:}\\
The drive work cost is
\begin{equation}
    W_{\text{dr}}=\Delta E_{S}+\Delta E_{A} + \Delta E_B, 
\end{equation}
where, unlike in the dephasing case, the exchange of energy with the bath must be accounted for in the drive work.

The entropy increase during the drive phase (with dissipation) is 
\begin{equation}
    \sigma_{\text{dr}}=\Delta S_{SA}-\beta Q_B \geq 0,
\end{equation}
where $\Delta S_{SA}=S\left[\sum_r p_r\hat{\rho}_{SA|r}\left(t_f\right)\right]-S_A\left(t_0\right)-S_S\left(t_0\right)$, and $t_f$ is the time at which the ancilla is readout.

\underline{Ancilla and memory reset:}\\
The work associated with resetting the memory and the ancilla is
\begin{eqnarray}
    W^{A}_{\text{reset}}&\geq& -\Delta E_A + \frac{1}{\beta}\Delta \mean{S_A} \\ \nonumber
    W^{M}_{\text{reset}}&\geq&  \frac{1}{\beta} H\left(p_r\right), 
\end{eqnarray}
where, $\Delta S_A=\sum_r p_r S\left[\hat{\rho}_{A|r}\left(t_f\right)\right]-S_A\left(t_0\right)$, and,
$\sum_r p_rS[\hat{\rho}_{A|r}(t_f)]=H\left(p_n\right)+\sum_n p_n H\left(p_{r|n}\right)-H\left(p_r\right)$. As discussed in Section~\ref{section:energetics}, if the coarse-graining subspaces $\hat{\Pi}_r$ are orthogonal, the term $\sum_n p_n H\left(p_{r|n}\right)$ vanishes. This scenario will be assumed for the remainder of the appendix.

The entropy production during the reset process is 
\begin{equation}
    \sigma_{\text{reset}}=-\Delta \mean{S_A}+H\left(p_r\right)\geq 0.
\end{equation}
The total entropy production is 
\begin{equation}
    \sigma_{\text{total}}=H\left(p_r\right)-\beta Q_B-\Delta \mean{S_A}+\Delta S_{SA}\geq 0
\end{equation}\\
\underline{Final balance:}\\
The change in bath energy corresponds to the heat exchange
\begin{equation}
    Q_B=-\Delta E_B . 
\end{equation}
Thus, the energy exchange with the bath can be expressed in terms of entropy exchange as
\begin{equation}
   \Delta E_B \geq -\frac{1}{\beta} \Delta S_{SA}.
\end{equation}
The overall work balance of the process is
\begin{equation}
  W_{\text{dr}}+ W^{A}_{\text{reset}}+ W^{M}_{\text{reset}} \geq \Delta E_{S} + \frac{1}{\beta}\Delta S_A + \frac{1}{\beta} H\left(p_r\right) -\frac{1}{\beta} \Delta S_{SA}.
\end{equation}
\section{Dynamical evolution} \label{appendix:dynamical_evolution}
In this appendix, we present detailed calculations of the qubit-cavity system dynamics, elaborating on the model and protocol outlined in sections~\ref{section:energetics} and~\ref{section:qubit-cavity-model}.

The initial qubit state, from which information is to be extracted, can be any two-level quantum state. This state can be represented using a density operator as:
\begin{equation}
    \hat{\rho}_S\left(t_0\right)=\begin{pmatrix}p_{g0} & \rho_{{ge}{0}} \\
    \rho_{{eg}{0}} & p_{e0}\end{pmatrix},
\end{equation}
where the probabilities must satisfy the normalization condition $p_{g0}+p_{e0}=1$, and the operator must be Hermitian, ensuring $\rho_{{eg}0}=\rho_{{ge}0}^{*}$.
The harmonic oscillator is initialized in its vacuum state
\begin{equation}
    \hat{\rho}_A\left(t_0\right)=\ket{0}\bra{0}.
\end{equation}

The interaction Hamiltonian between the system and the ancilla is provided in Eq.~\ref{eq:hamiltonian_system-ancilla}. During this interaction, the Hamiltonian remains diagonal in the qubit basis. Assuming the following basis, 
\begin{equation}
    \ket{g}=\begin{pmatrix}
        1 \\
        0
    \end{pmatrix}\hspace{1 cm}
    \ket{e}=\begin{pmatrix}
            0\\
            1
        \end{pmatrix},
\end{equation}
the Hamiltonian can be expressed as
\begin{equation}
    \hat{H}_{SA}\left(t\right)=\begin{pmatrix}
        \frac{\omega}{2}\hat{I}+\omega_a\hat{a}^{\dagger}\hat{a}+\mu\left( t\right)\left(\hat{a}+\hat{a}^{\dagger}\right) & 0 \\
        0 & -\frac{\omega}{2}\hat{I}+\omega_a\hat{a}^{\dagger}\hat{a}
    \end{pmatrix}.
\end{equation}

The evolution is most effectively analyzed in the interaction picture, where the density operator and the time-dependent component of the Hamiltonian are represented as:
\begin{equation}
\hat{\rho}_I\left(t_1\right)=\hat{U}_0^{\dagger}\left(\Delta t\right)\hat{\rho}_S\left(t_0\right)\hat{U}_0\left(\Delta t\right) \hspace{1 cm} \hat{V}_I\left(\Delta t\right)=\hat{U}_0\left(\Delta t\right)^{\dagger}\hat{V}\left(\Delta t\right)\hat{U}_0\left(\Delta t\right), 
\end{equation}
where $\Delta t=t_1-t_0$ and $\hat{U}_0\left(\Delta t\right)=e^{-i\hat{H}_0\Delta t}$.

In the interaction picture, the von Neumann equation takes the form:
\begin{equation}
    \frac{\partial\hat{\rho}_I}{\partial t'}=-i\left[\hat{V}_I\left(t'\right),\hat{\rho}_I\left(t'\right)\right],
\end{equation}
where $\hbar=1$. The evolution of the density operator in the interaction picture is given by:
\begin{equation}
\label{eq:equation_density_operator}
    \hat{\rho}_I\left(t_1\right)=\hat{\rho}_I\left(t_0\right)-i\int_{t_0}^{t_1} \left[\hat{V}_I\left(t'\right),\hat{\rho}_I\left(t'\right)\right]dt'.
\end{equation}

In the qubit basis, $\hat{V}_I\left( t'\right)$ is:
\begin{eqnarray}
    \hat{V}_I\left(t'\right)&=&\begin{pmatrix}
        e^{i\left(\omega\hat{I}+\omega_A\hat{a}^{\dagger}\hat{a}\right)t'}\mu\left( t'\right)\left(\hat{a}+\hat{a}^{\dagger}\right)e^{-i\left(\omega\hat{I}+\omega_A\hat{a}^{\dagger}\hat{a}\right)t'} & 0 \\
        0 & 0
    \end{pmatrix}\\&=& 
    \begin{pmatrix}
        \mu\left(t'\right)\left(\hat{a}^{\dagger}e^{i\omega_a t'}+\hat{a}e^{-i\omega_a t'}\right) & 0\\
        0 & 0
    \end{pmatrix}.
\end{eqnarray}

The formal solution to this equation is
\begin{equation}
    \hat{\rho}_I\left(t_1\right)=\hat{U}_I\left(\Delta t\right)\hat{\rho}_I\left(t_0\right)\hat{U}_I^{\dagger}\left(\Delta t\right),
\end{equation}
where, 
\begin{equation}
    \hat{U}_I\left(\Delta t\right)=\mathcal{T}\left(\text{exp}\left(-i\int_{t_0}^{t_1}\hat{V}_I\left(t'\right)dt'\right)\right).
\end{equation}

The only relevant component of $\hat{V}_I\left(\Delta t\right)$ is the matrix's ground state. Therefore, the operator's element is:
\begin{equation}
U_{I,g}\left(\Delta t\right)=\mathcal{T}\left(\text{exp}\left(-i\int_{t_0}^{t_1}\mu\left(t'\right)\left(\hat{a}^{\dagger}e^{i\omega_at'}+\hat{a}e^{-i\omega_at'}\right)dt'\right)\right),
\end{equation}
where the integral cannot be solved directly because the operator does not commute with itself at different times. However, if we assume that $\mu\left(t'\right)\in\mathbb{R}$, the operator can be expressed as a displacement operator, given by:
\begin{equation}
D\left(\alpha\right)=\text{exp}\left(-i\mu\left(t'\right)\left(\hat{a}^{\dagger}e^{i\omega_at'}+\hat{a}e^{-i\omega_at'}\right)\right),
\end{equation}
with $\alpha_1=-i\mu\left(t'\right)e^{i\omega_at'}$. The operator $U_{I,g}\left(\Delta t\right)$ can be computed as the product of an infinite sequence of unitary operators, expressed as:
\begin{equation}
   \hat{U}_I\left(\Delta t\right)=\lim_{n\to \infty}\prod_{j=0}^{n}\text{exp}\left(-i\mu\left(j\frac{\Delta t}{n}\right)\left(\hat{a}^{\dagger}e^{i\omega_aj\frac{\Delta t}{n}}+\hat{a}e^{-i\omega_aj\frac{\Delta t}{n}}\right)\frac{\Delta t}{n}\right).
\end{equation}

Using the property:
\begin{equation}
\hat{D}\left(\alpha\right)\hat{D}\left(\beta\right)=e^{\frac{\left(\alpha\beta^{*}-\alpha^{*}\beta\right)}{2}}\hat{D}\left(\alpha+\beta\right),
\end{equation}
the unitary operator can be expressed in the following form:
\begin{eqnarray}
   \hat{U}_I\left(\Delta t\right)&=& \lim_{n\to \infty}\text{exp}\left[\sum_{l=1}^{n-1}i\sin{\frac{\omega_a\Delta tl}{n}}\sum_{b=1}^{n-l+1}\prod_{c=0}^{b}\mu\left(\frac{c \Delta t}{n}\right)\frac{\Delta t}{n}\right]\\ \nonumber
   &\times& \text{exp}\left[\sum_{j=0}^n -i\mu\left(j\frac{\Delta t}{n}\right)\left(\hat{a}^{\dagger}e^{i\omega_aj\frac{\Delta t}{n}}+\hat{a}e^{-i\omega_aj\frac{\Delta t}{n}}\right)\frac{\Delta t}{n}\right]\\ \nonumber
   &=&e^{-i\xi}\text{exp}\left(\int_{t_0}^{t_1}-i\mu\left(t'\right)\left(\hat{a}^{\dagger}e^{i\omega_at'}+\hat{a}e^{-i\omega_at'}\right)dt'\right)=e^{i\beta}\hat{D}\left(\alpha\right), 
\end{eqnarray}
where, 
\begin{equation}
    \alpha=-i\int_{t_0}^{t_1}\mu\left(t'\right)e^{i\omega_at'}dt'.
\end{equation}

Following this, the full state of the system is:
\begin{eqnarray}
\rho_{gg}\left(t_1\right)&=&p_{g_0}e^{-|\alpha|^2}\sum_{n=0}^{\infty}\sum_{m=0}^{\infty}\frac{\alpha^n}{\sqrt{n!}}\frac{\alpha^{*m}}{\sqrt{m!}}\ket{n}\bra{m} \\ \nonumber
\rho_{ge}\left(t_1\right)&=&\rho_{{ge}_{0}}e^{-2i\omega t_1}e^{\frac{-|\alpha|^2}{2}}e^{-i\xi}\sum_{n=0}^{\infty}e^{i\omega_a t_1 n}\frac{\alpha^n}{\sqrt{n!}}\ket{n}\bra{0}\\ \nonumber
\rho_{eg}\left(t_1\right)&=&\rho_{{eg}_{0}}e^{2i\omega t_1}e^{\frac{-|\alpha|^2}{2}}e^{i\xi}\sum_{n=0}^{\infty}e^{i\omega^{*}_a t_1n}\frac{\alpha^n}{\sqrt{n!}}\ket{0}\bra{n}\\ \nonumber
\rho_{ee}\left(t_1\right)&=&p_{{e}_0}\ket{0}\bra{0}.
\end{eqnarray}

However, if we measure the harmonic oscillator state after dephasing, the measurement would be semi-strong; detecting at least one photon would reveal that the system is in the ground state. To ensure a fully weak measurement, a second displacement operator, $\hat{D}_2\left(\alpha_2\right)$, is applied to the harmonic oscillator over a time interval $\Delta t_2$. The state of the entire system after this second displacement becomes:
\begin{eqnarray}
\rho_{gg}\left(t_2\right)&=&p_{g_0}e^{-|\alpha_1|^2}\sum_{n=0}^{\infty}\sum_{m=0}^{\infty}e^{i\omega_at_2n}e^{-i\omega_at_2
m}\frac{\alpha_1^n}{\sqrt{n!}}\frac{\alpha_1^{*m}}{\sqrt{m!}}\ket{n}\bra{m} \\ \nonumber
\rho_{ge}\left(t_2\right)&=&\rho_{{ge}_{0}}e^{-i\phi}e^{\frac{-\left(|\alpha_1|^2+|\alpha_2|^2\right)}{2}}\sum_{n=0}^{\infty}\sum_{m=0}^{\infty}e^{i\omega_a t_2 n}e^{-i\omega_a t_2m}\frac{\alpha_1^n}{\sqrt{n!}}\frac{\alpha_2^{*m}}{\sqrt{m!}}\ket{n}\bra{m}\\ \nonumber
\rho_{eg}\left(t_2\right)&=&\rho_{{eg}_{0}}e^{i\phi}e^{\frac{-\left(|\alpha_1|^2+|\alpha_2|^2\right)}{2}} \sum_{n=0}^{\infty}\sum_{m=0}^{\infty}e^{-i\omega_at_2n}e^{i\omega_at_2m}\frac{\alpha_1^{*n}}{\sqrt{n!}}\frac{\alpha_2^m}{\sqrt{m!}}\ket{m}\bra{n}\\ \nonumber
\rho_{ee}\left(t_2\right)&=&p_{e_0}e^{-|\alpha_2|^2}\sum_{n=0}^{\infty}\sum_{m=0}^{\infty}e^{i\omega_at_2n}e^{-i\omega_at_2m}\frac{\alpha_2^n}{\sqrt{n!}}\frac{\alpha_2^{*m}}{\sqrt{m!}}\ket{n}\bra{m},
\end{eqnarray}
where $\alpha_1=\alpha+\alpha_2$, and $\phi$ is the total phase acquired by the unitary process, including the displacement on the ancilla.

After the system-ancilla interaction and the second displacement, the harmonic oscillator undergoes dephasing for a sufficiently long period to ensure complete dephasing. During this interval, the system $S$ and the ancilla $A$ remain non-interacting. At the end of this process, the state of the combined system and ancilla is given by Eq.~\ref{eq:full_system_state_after_dephasing}.

To extract information about the system state, we might measure the number of excitations in the cavity, represented by the operator $\ket{n}\bra{n}$. The conditional state of the system, given that $n$ excitations are detected in the cavity, is described by Eq.~\ref{eq:conditional_system_state}. The conditional state of the harmonic oscillator is
\begin{equation}
    \rho_{{A|n}_{1,1}}\left(t_f\right)=\ket{n}\bra{n}. 
\end{equation}

The conditional system-ancilla state is
\begin{eqnarray}
p_{g|n}\left(t_f\right)&=&\frac{p_{g_0}}{p_n}e^{-|\alpha_1|^2}\frac{|\alpha_1|^{2n}}{n!}\ket{n}\bra{n} \\ \nonumber
\rho_{ge|n}\left(t_f\right)&=&\frac{\rho_{{ge}_{0}}}{p_n}e^{-\phi}e^{\frac{-\left(|\alpha_1|^2+|\alpha_2|^2\right)}{2}}\frac{\alpha_1^n}{\sqrt{n!}}\frac{\alpha_2^{*n}}{\sqrt{n!}}\ket{n}\bra{n}\\ \nonumber
\rho_{eg|n}\left(t_f\right)&=&\frac{\rho_{{eg}_{0}}}{p_n}e^{i\phi}e^{\frac{-\left(|\alpha_1|^2+|\alpha_2|^2\right)}{2}}\frac{\alpha_1^{*n}}{\sqrt{n!}}\frac{\alpha_2^n}{\sqrt{n!}}\ket{n}\bra{n}\\ \nonumber
p_{e|n}\left(t_f\right)&=&\frac{p_{e_0}}{p_n}e^{-|\alpha_2|^2}\frac{|\alpha_2|^{2n}}{n!}\ket{n}\bra{n},
\end{eqnarray}
where the probability to get $n$ is $p_n=p_{g_0}e^{-|\alpha_1|^2}\frac{|\alpha_1|^{2n}}{n!}+p_{e_0}e^{-|\alpha_2|^2}\frac{|\alpha_2|^{2n}}{n!}$.

To perform a weak measurement, we assume that the two displacements are nearly identical, as specified in Eq.~\ref{eq:alpha_weak_measurement}. Under this assumption, only terms linear in $\epsilon$ are significant. Consequently, Eq.~\ref{eq:full_system_state_after_dephasing} simplifies to:
\begin{eqnarray}
\rho_{gg}\left(t_f\right)&\approx& p_{g_0}e^{-|\overline{\alpha}|^2}\sum_{n=0}^{\infty}\frac{|\overline{\alpha}|^{2n}}{n!}\left(1+2\epsilon \text{Re}\left(\overline{\alpha}\right)\left(\frac{n}{|\overline{\alpha}|^2}-1\right)\right)\ket{n}\bra{n} \\ \nonumber
\rho_{ge}\left(t_f\right)&\approx& \rho_{{ge}_{0}}e^{-i\phi}e^{-|\overline{\alpha}|^2}\sum_{n=0}^{\infty}\frac{|\overline{\alpha}|^{2n}}{n!}\left(1-\frac{2ni\epsilon\text{Im}\left(\overline{\alpha}\right)}{|\overline{\alpha}|^2}\right)\ket{n}\bra{n} \\ \nonumber
\rho_{eg}\left(t_f\right)&\approx& \rho_{{eg}_{0}}e^{i\phi}e^{-|\overline{\alpha}|^2}\sum_{n=0}^{\infty}\frac{|\overline{\alpha}|^{2n}}{n!}\left(1+\frac{2ni\epsilon\text{Im}\left(\overline{\alpha}\right)}{|\overline{\alpha}|^2}\right)\ket{n}\bra{n} \\ \nonumber
\rho_{ee}\left(t_f\right)&\approx& p_{e_0}e^{-|\overline{\alpha}|^2}\sum_{n=0}^{\infty}\frac{|\overline{\alpha}|^{2n}}{n!}\left(1+2\epsilon \text{Re}\left(\overline{\alpha}\right)\left(1-\frac{n}{|\overline{\alpha}|^2}\right)\right)\ket{n}\bra{n}.  
\end{eqnarray}

The conditional state of the system, given that $n$ excitations were measured, is expressed as:
\begin{eqnarray}
    p_{g|n}\left(t_f\right)&\approx& \frac{p_{g_0}}{p_{\epsilon n}}e^{-|\overline{\alpha}|^2}\frac{|\overline{\alpha}|^{2n}}{n!}\left(1+2\epsilon \text{Re}\left(\overline{\alpha}\right)\left(\frac{n}{|\overline{\alpha}|^2}-1\right)\right) \\ \nonumber
    \rho_{ge|n}\left(t_f\right)&\approx& \frac{\rho_{{ge}_0}}{p_{\epsilon n}}e^{-i\phi}e^{-|\overline{\alpha}|^2}\frac{|\overline{\alpha}|^{2n}}{n!}\left(1-\frac{2ni\epsilon\text{Im}\left(\overline{\alpha}\right)}{|\overline{\alpha}|^2}\right) \\ \nonumber
    \rho_{eg|n}\left(t_f\right)&\approx& \frac{\rho_{{eg}_0}}{p_{\epsilon n}}e^{i\phi}e^{-|\overline{\alpha}|^2}\frac{|\overline{\alpha}|^{2n}}{n!}\left(1+\frac{2ni\epsilon\text{Im}\left(\overline{\alpha}\right)}{|\overline{\alpha}|^2}\right)\\ \nonumber
    p_{e|n}\left(t_f\right)&\approx& \frac{p_{e_0}}{p_{\epsilon n}}e^{-|\overline{\alpha}|^2}\frac{|\overline{\alpha}|^{2n}}{n!}\left(1+2\epsilon \text{Re}\left(\overline{\alpha}\right)\left(1-\frac{n}{|\overline{\alpha}|^2}\right)\right),
    \end{eqnarray}
where the probability of measuring each value $n$ is given by:
\begin{equation}
    p_{\epsilon n}=e^{-|\overline{\alpha}|^2}\frac{|\overline{\alpha}|^{2n}}{n!}\left(1+2\epsilon\text{Re}\left(\overline{\alpha}\right)\left(p_{g0}-p_{e0}\right)\left(\frac{n}{|\overline{\alpha}|^2}-1\right)\right).
\end{equation}
\section{Initial thermal ancilla state}\label{appendix:initial_thermal_state}
In this appendix, we examine the performance of the qubit measurement when the ancilla is initialized in a thermal state of finite temperature. The protocol, as described in section~\ref{section:qubit-cavity-model}, is modified by replacing the initial ancilla vacuum state with a thermal state characterized by $\beta \omega_B = 3$, where $\hbar = 1$. This value was chosen for being of the correct order of magnitude to describe, e.g. superconducting qubit readout via their coupling to microwave cavities \cite{ficheux2018dynamics}. After applying the protocol, we evaluate the resulting performance metrics and analyze the associated work bound.

\begin{figure}
    \centering
    \includegraphics[width=0.5\linewidth]{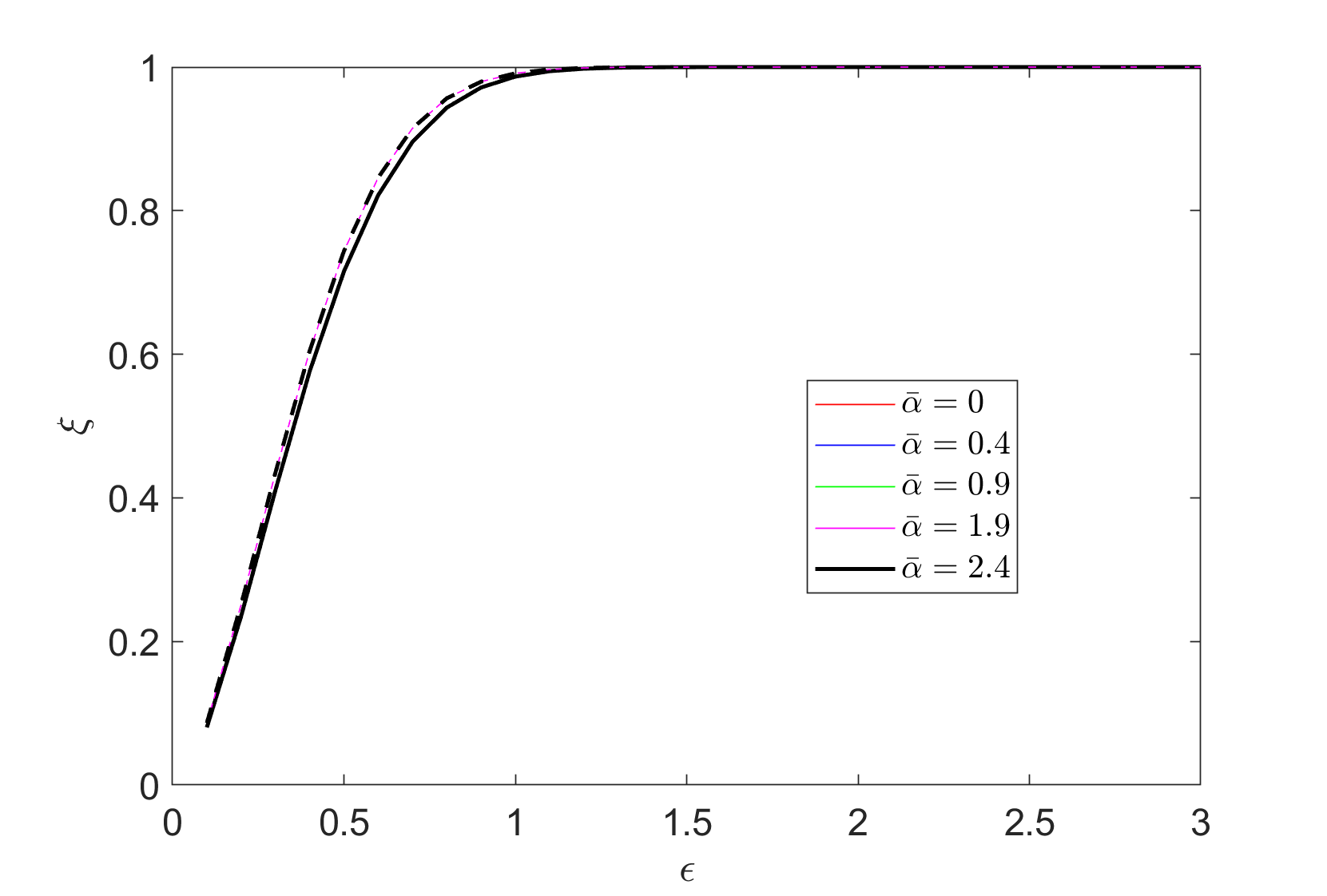}
    \caption{Strength as a function of the parameter $\epsilon$, shown for various values of $\bar{\alpha}$. The plot compares the strength for an initial vacuum ancilla state (solid line) and a thermal ancilla state (dashed line). The initial qubit state is $\frac{1}{\sqrt{2}}\left(\ket{g}+\ket{e}\right)$}.
    \label{fig:strength_thermal_initial_state}
\end{figure}

Fig.~\ref{fig:strength_thermal_initial_state} shows the strength, as defined in Eq.~\ref{eq:strength_measure}, plotted in terms of $\epsilon$ for various values of $\bar{\alpha}$, having $\ket{\text{ref}}=\frac{1}{\sqrt{2}}\left(\ket{g}+\ket{e}\right)$ as the initial qubit state. As expected, and consistent with the case of an initial vacuum state, the strength is independent of $\bar{\alpha}$. Notably, for the same value of $\epsilon$, the strength is higher when the ancilla begins in a thermal state (dashed line) compared to when it starts in a vacuum state (solid line). This is due to an increased magnitude of the dephasing experienced by the backaction.

\begin{figure}
    \centering
    \includegraphics[width=0.5\linewidth]{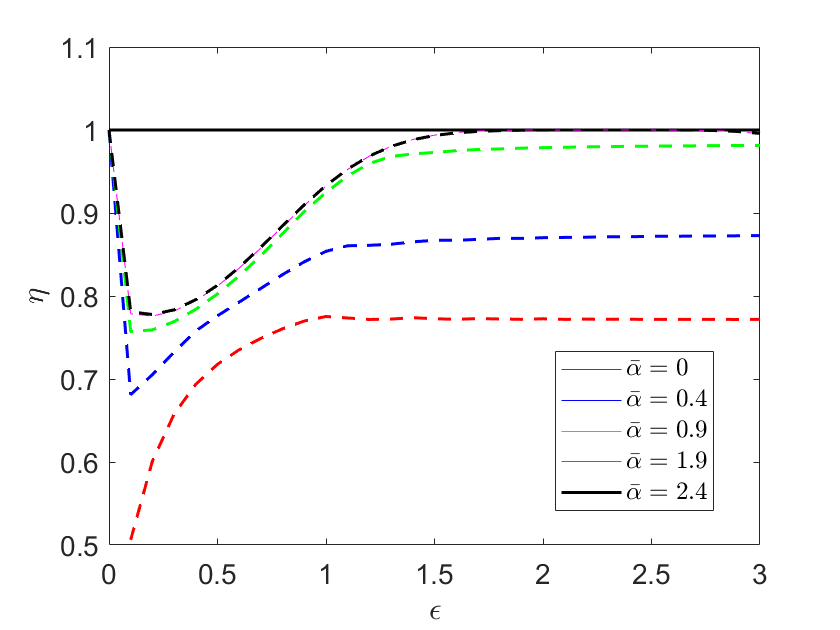}
    \caption{Measure of efficiency $\eta$ as a function of $\epsilon$ for various values of $\bar{\alpha}$. The solid line represents the initial vacuum ancilla state ($\eta=1$ whatever $\bar\alpha$), while the dashed lines represent the initial thermal ancilla state. The initial qubit state is $\frac{1}{\sqrt{2}}\left(\ket{g}+\ket{e}\right)$.}
    \label{fig:efficiency_thermal_initial_state}
\end{figure}

Fig.~\ref{fig:efficiency_thermal_initial_state} shows the detection efficiency, $\eta$, as defined in Eq.~\ref{eq:efficiency_mutual_information}, plotted in terms of $\epsilon$ for various values of $\bar{\alpha}$. The results are presented for an initial vacuum ancilla state (solid line) and a thermal ancilla state (dashed lines). Without coarse-graining, the efficiency with an initial zero-temperature ancilla state remains constant at $1$ for all values of $\bar{\alpha}$. In contrast, with the initial thermal state, the efficiency depends on both $\epsilon$ and $\bar{\alpha}$. For small values of $\epsilon$ (weak measurements), the efficiency is low, decreasing further for smaller $\bar{\alpha}$. For larger $\epsilon$ (strong measurements), the efficiency approaches a plateau whose value depends on $\bar{\alpha}$. When $\bar{\alpha}$ is sufficiently large, the efficiency ultimately reaches $1$.

\begin{figure}
    \centering    \includegraphics[width=0.5\linewidth]{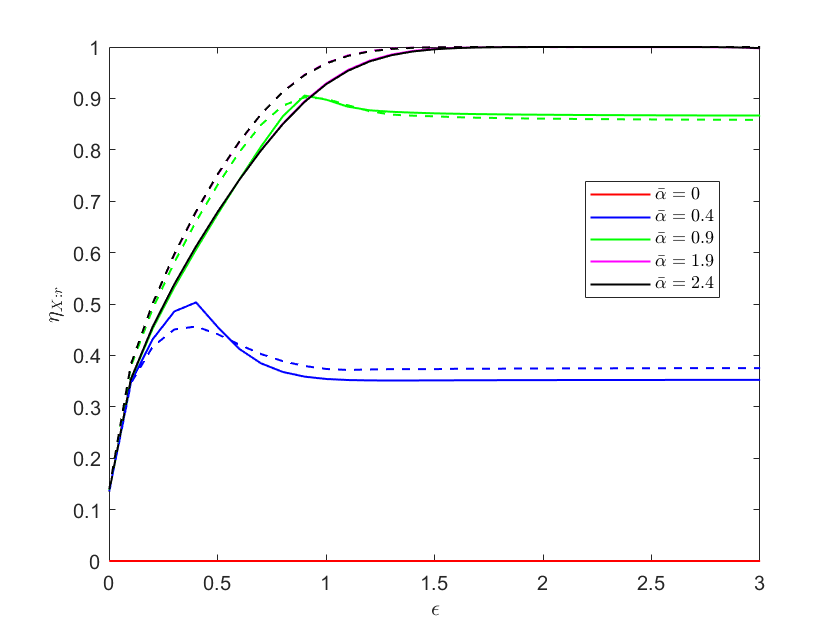}
    \caption{Normalized classical mutual information efficiency $\eta_{X:r}$ as a function of $\epsilon$ for various values of $\bar{\alpha}$. The plot shows results for the initial vacuum ancilla state (solid lines) and the initial thermal ancilla state (dashed lines). The initial qubit state is $\frac{1}{\sqrt{2}}\left(\ket{g}+\ket{e}\right)$.}
    \label{fig:eta_I_thermal_initial_ancilla_state}
\end{figure}

In Fig.~\ref{fig:eta_I_thermal_initial_ancilla_state}, the normalized classical mutual information, $\eta_{X:r}$, as defined in Eq.~\ref{eq:efficiency_mutual_information}, is plotted as a function of $\epsilon$ for various values of $\bar{\alpha}$. Results are shown for both a vacuum initial ancilla state (solid lines) and a thermal initial ancilla state (dashed lines). Although the overall behavior is similar in both cases, the normalized classical mutual information is larger for either the vacuum or thermal state, depending on the values of of $\bar{\alpha}$ and $\epsilon$. We note that the (un-normalized) classical mutual information is consistently larger for the vacuum state than for the thermal state. The regimes where $\eta_{X:r}$ is larger at higher temperature are due to a decrease of the normalization factor (i.e. the Holevo information). When considering the overall performance $\xi \eta \eta_{X:r}$, the protocol using a vacuum initial ancilla state is consistently better than the one using a thermal state.

\begin{figure}
    \centering
    \includegraphics[width=0.5\linewidth]{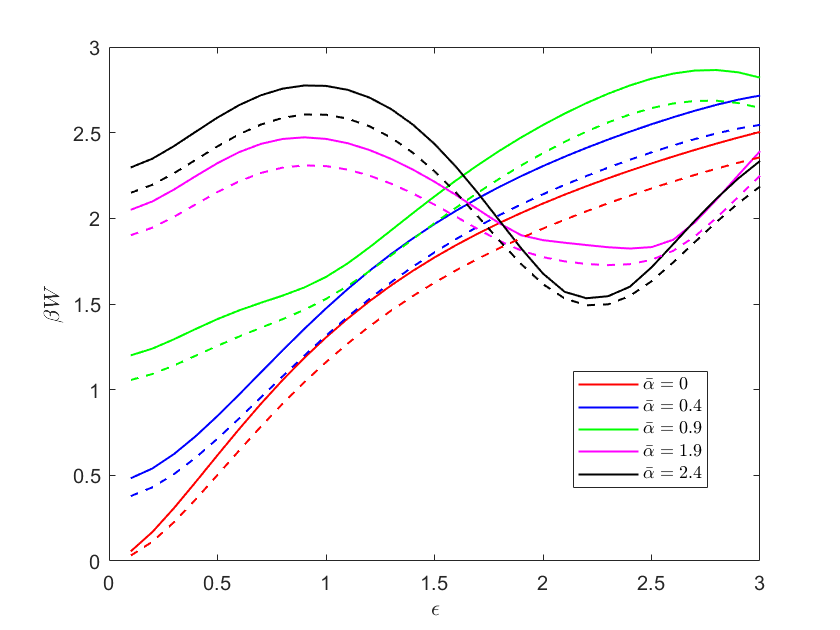}
    \caption{Work bound as a function of $\epsilon$ for various values of $\bar{\alpha}$. The plot includes results for both the initial vacuum ancilla state (solid lines) and the initial thermal ancilla state (dashed lines). The initial qubit state is $\frac{1}{\sqrt{2}}\left(\ket{g}+\ket{e}\right)$.}
    \label{fig:work_thermal_initial_ancilla_state}
\end{figure}

In Fig.~\ref{fig:work_thermal_initial_ancilla_state}, we plot the work bound of the protocol, as given by Eq.~\ref{eq:work_bound_dephasing}. For the same values of $\bar{\alpha}$ and $\epsilon$, the work bound is higher when the initial ancilla state is thermal compared with the vacuum. This result is in agreement with observations in Ref.\cite{latune2024thermodynamically} that the cost increases with the amount of extracted information and the measurement efficiency, which both decrease with the temperature. Indeed, when starting with a thermal state, we have:
\be
p_n = \sum_m p_m^\text{th}\left(p_{g0}|\bra{n}\hat D(\alpha_1)\ket{m}|^2+p_{e0}|\bra{n}\hat D(\alpha_2)\ket{m}|^2\right), \quad\quad\text{with}\quad p_m^\text{th}=(1-e^{-\beta\hbar\omega_a})e^{-m\beta\hbar\omega_a}.
\ee
In the limit of extremely high temperature $\beta\to 0$, the distribution $p_m^\text{th}$ becomes approximately constant over the support of $|\bra{n}\hat D(\alpha_1)\ket{m}|^2$ (considered as a function of $m$), such that $p_n \to p_n^\text{th}$. As a consequence, the result of the measurement is not correlated anymore to the value of the measured observable, and $H(p_n)=S_A(t_0)$, leading to a vanishing work cost.

In summary, when using an initial thermal ancilla state, for a given value of $\epsilon$ and $\bar{\alpha}$, the information extracted from the measurement protocol is reduced, but the work bound is also lower.

\section{comparative analysis of work costs in continuous weak measurements versus strong measurements}\label{appendix:general_strong_versus_weak_measurement_cost}
In this appendix, we present a comprehensive analysis of the work cost associated with a single strong measurement and continuous weak measurements. First, we examine the scaling behavior of the work cost. Next, we analyze the scaling of dumping in the case of weak measurements. Finally, we compare the work costs of both approaches based on the previously derived scaling relationships.

\subsection{Scaling of the work cost}\label{appendix_subsection:scaling_of_the_work_cost}

We start from Eq.~\eqref{eq:work_bound_dephasing}, in the case of a measurement of an observable commuting with the system Hamiltonian, such that $\Delta E_S=0$, and the work lower bound is given by the Shannon entropy $H(p_n)$ of the ancilla eigenstate distribution $p_n$. We then use phenomenological assumptions on distribution $p_n$ to draw conclusion on the work bound.

We assume that after the interaction with the system (characterized by an interaction strength $\epsilon$) and the pure dephasing steps, the populations of the energy eigenstates of the ancilla obey:
\be  
p_n(\epsilon) = \sum_k p_k G_{u_k(\epsilon)}^{\sigma_k(\epsilon)}(n),
\ee
where $p_k$ are the probabilities of the different eigenstates of the measured system observable (i.e. the measurement result in an ideal projective measurement), and the functions $G_{u_k(\epsilon)}^{\sigma_k(\epsilon)}(n)$ describe distributions of energy eigenstate of the ancilla conditioned to the system being in the $k$th eigenstate. We assume these functions are smooth, single-peaked, of average $u_k(\epsilon)$ and standard deviation $\sigma_k(\epsilon)$. In the limit $\epsilon\to 0$, where no interaction takes place, we assume $\underset{\epsilon\to 0}{\text{lim}}\, u_k(\epsilon) =0$, and $\underset{\epsilon\to 0}{\text{lim}}\, \sigma_k(\epsilon) =\sigma_0$. Conversely, the limit $\epsilon\to \infty$ corresponds to an unambiguous imprint of the measurement results, that is $\vert u_k-u_k'\vert \to \infty$, $\forall k\neq k'$, and for sufficiently large $\epsilon$, we have $\sigma_k(\epsilon),\;\sigma_{k'}(\epsilon) \geq \vert u_k-u_k'\vert$.\\

Those conditions, which quite generally correspond to a behavior of pointer for a measurement whose strength is controlled by $\epsilon$, strongly constrain the behavior of the Shannon entropy of the distribution. More precisely, we have in the weak measurement limit:
\be  
H(p_n(\epsilon)) \underset{\epsilon\to 0}{\rightarrow} H\left( G_{0}^{\sigma_0}(n)\right) \equiv H_0.
\ee
In the strong measurement limit, as the functions $G_{u_k(\epsilon)}^{\sigma_k(\epsilon)}$ have disjoint supports for different $k$, we have:
\be
H(p_n(\epsilon)) \underset{\epsilon\to \infty}{\sim} H(p_k)+\sum_k p_k H\left( G_{u_k(\epsilon)}^{\sigma_k(\epsilon)}(n)\right).
\ee

In the most simple case, the interaction only affects the average of the pointer distributions, that is $\sigma_k(\epsilon)=\sigma_0$. As the Shannon entropy of a single-peak distribution typically scales as the log of its standard deviation, the work cost interpolates between two plateaus $H_0$ and $H_0+H(p_k)$ when $\epsilon$ is varied from $0$ to $\infty$. Importantly, in this case, the work cost saturates at very large $\epsilon$.\\
However, physical interactions like the Jayne-Cummings interaction tend to increase the variance while shifting the average number of excitation of the ancilla, leading to $\sigma_k(\epsilon)\to \infty$ when $\epsilon\to \infty$, leading to a work cost which keeps increasing with large $\epsilon$.\\

In the specific example treated in this article, we can identify the functions $G_{u_k(\epsilon)}^{\sigma_k(\epsilon)}(n) \equiv e^{-|\alpha_k|^2}\frac{|\alpha_k|^{2n}}{n!}$ with $\alpha_e=\bar\alpha-\epsilon$ and $\alpha_g=\bar\alpha+\epsilon$. Thus, $u_{e/g}(\epsilon)=|\bar\alpha\mp\epsilon|^2$ and $\sigma_{e/g}(\epsilon) = |\bar\alpha\mp\epsilon|$. For a fixed $\bar{\alpha}$, when increasing $\epsilon$ from $0$ to $\infty$, we can identify three regimes. For $\epsilon \ll 1$, $H(p_n)$ is of the order of the entropy of the Shannon entropy of the Poisson distribution $e^{-|\bar\alpha|^2}\frac{|\bar\alpha|^{2n}}{n!}$. When $\epsilon\gg\bar\alpha$, $H(p_n)$ is of the order of $H(p_k)+ H\left(e^{-|\epsilon|^2}\frac{|\epsilon|^{2n}}{n!}\right)$, which increases with $\epsilon$ due to the second term. In between, the variance $\sigma_e$ turns out to have a non monotonic behavior. In particular, when $\epsilon\simeq \bar\alpha$, the variance of the distribution associated to $k=e$ vanishes, and consequently so does $H(G_{u_e(\epsilon)}^{\sigma_e(\epsilon)}(n))$. This property may lead to a non-monotonic behavior for the work cost for $\epsilon \in [0,\bar\alpha]$. For instance, if $\bar\alpha \gtrsim 1$, the two distributions $G_{u_k(\epsilon)}^{\sigma_k(\epsilon)}(n)$ are already orthogonal when $\epsilon=\bar\alpha$, and $H(p_n)\sim H(p_k)+p_g H\left(e^{-|2\bar\alpha|^2}\frac{|2\bar\alpha|^{2n}}{n!}\right)$. Moreover, $H\left(e^{-|x|^2}\frac{|x|^{2n}}{n!}\right)$ is sublinear in $x$ as long as $x >1$. Those facts can lead to situations where the minimum work cost can decrease with $\epsilon$ in the interval $[0,\bar\alpha]$, or even reach its overall minimum at $\epsilon=\bar\alpha$ (see Fig.~\ref{fig:work_in_terms_of_epsilon_for_different_alpha}).

\subsection{Weak measurement}\label{appendix:weak_measurement}

On general grounds, one can model the interaction step as a conditioned unitary on the system and ancilla:
\begin{eqnarray}
    \hat\rho_{SA}(t_2) = \sum_{k,k'} \hat U_k \hat U_u \hat\rho_A(0) \hat{U}_u^\dagger \hat{U}_{k'}^\dagger \otimes \hat\pi_k\hat\rho_S(t_1)\hat\pi_{k'},
\end{eqnarray}
where we have singled out the unconditionned part of the ancilla unitary evolution $\hat{U}_u$. The resulting unconditioned post-measurement system state is:
\begin{eqnarray}
    \hat\rho_{S}(t_2) = \sum_{k,k'} \text{Tr}\{\hat U_{k'}^\dagger \hat U_k \hat U_u \hat\rho_A(0) \hat U_u^\dagger \} \hat\pi_k\hat\rho_S(t_1)\hat\pi_{k'},
\end{eqnarray}

A weak measurement is achieved in the limit where $\hat{U}_k$ is close to the identity. More precisely, an interaction Hamiltonian of the form $\hat V = \mu \sum_k \hat V_k \otimes \hat\pi_k$ implies for sufficiently short interaction time $\mu\tau = \epsilon \ll 1$:
\begin{eqnarray}
    \hat U_k = \hat\idop -i \epsilon \hat V_k - \frac{\epsilon^2}{2} \hat V_k^2 + {\cal O}(\epsilon^3),
\end{eqnarray}
and then
\begin{eqnarray}
    \hat U_{k'}^\dagger \hat U_k &=& \hat\idop -i \epsilon (\hat V_k-\hat V_{k'}) - \epsilon^2 \left(\hat V_{k'} \hat V_k-\frac{1}{2}\hat V_k^2-\frac{1}{2}\hat V_{k'}^2\right)+ {\cal O}(\epsilon^3).
\end{eqnarray}
We therefore obtain:
\begin{eqnarray}
    \hat\rho_{S}(t_2) = \sum_{k,k'} \left(1-i\epsilon\mean{\hat V_k-\hat V_{k'}}- \epsilon^2 \left(\mean{\hat V_{k'} \hat V_k}-\frac{1}{2}\mean{\hat V_k^2}-\frac{1}{2}\mean{\hat V_{k'}^2}\right)\right)\hat\pi_k\hat\rho_S(t_1)\hat\pi_{k'}+ {\cal O}(\epsilon^3),
\end{eqnarray}
where the averages are taken in the ancilla state $\hat U_u \hat\rho_A(0) \hat U_u^\dagger$. 
We rewrite:
\begin{align}
    \hat\rho_{S}(t_2) &= \sum_{k,k'} \hat\pi_k e^{-i\epsilon\hat V_\text{av}} \hat\rho_S(t_1)e^{i\epsilon\hat V_\text{av}} \hat\pi_{k'}- \epsilon^2\sum_{k,k'} \left(\mean{\hat V_{k'} \hat V_k}-\frac{1}{2}\mean{\hat V_k^2}-\frac{1}{2}\mean{\hat V_{k'}^2}\right)\hat\pi_k  \hat\rho_S(t_1) \hat\pi_{k'}+ {\cal O}(\epsilon^3)\nonumber\\
    &= e^{-i\epsilon\hat V_\text{av}} \hat\rho_S(t_1)e^{i\epsilon\hat V_\text{av}}  - \epsilon^2\sum_{k,k'}  \left(\mean{\hat V_{k'} \hat V_k}-\frac{1}{2}\mean{\hat V_k^2}-\frac{1}{2}\mean{\hat V_{k'}^2}\right)\hat\pi_k  \hat\rho_S(t_1) \hat\pi_{k'}+ {\cal O}(\epsilon^3) 
\end{align}
where $\hat V_\text{av}=\sum_k{\mean{\hat V_k}\hat\pi_k}$. From this equation, we can see that the coherence damping induced by the measurement process is characterized by a rate scaling as $\epsilon^2$. More precisely:
\begin{eqnarray}
    \frac{\Delta\rho_{kk'}}{\tau} = -\epsilon^2 \left(\mean{\hat V_{k'} \hat V_k}-\frac{1}{2}\mean{\hat V_k^2}-\frac{1}{2}\mean{\hat V_{k'}^2}\right)\rho_{kk'}
\end{eqnarray}
implies an expontential decay of the coherence amplitude.\\

\subsection{Work cost scaling}\label{appendix:trial_scaling_of_the_work_cost}
The Shannon entropy of the distribution for small $\epsilon$ is given by $H_0$. Since the measurement must be repeated a number of times proportional to $\frac{1}{\epsilon^2}$, the total work associated with concatenated weak measurements is therefore:
\begin{equation}
\label{eq:scaling_work_weak}
    W^{\text{weak}}\propto\frac{H_0}{\epsilon^2}
\end{equation}
On the other hand, in the case of a strong measurement, the probability distribution is expected to be well separated. For large $\epsilon$, let us assume that the distribution in $n$ consists of $K$ well-separated Gaussian components, corresponding to the eigenstates of the system. In this scenario, the total work is given by the sum of the Shannon entropies of all these Gaussian distributions.
\begin{equation}
    W^{\text{strong}}=K\left[\frac{1}{2}\log{\left(2\pi\sigma^2\right)}+\frac{1}{2}\right],
\end{equation}
where $\sigma$ represents the standard deviation of the Gaussian probability distribution. Suppose the standard deviation scales as $\epsilon^l$, where $l$ is a specific exponent determined by the particular model. In most physical cases, the exponent is either 0 or 1. In this case, the work associated with a strong measurement will scale with $\epsilon$ as follows:
\begin{equation}
\label{eq:scaling_work_strong}
    W^{\text{strong}}\propto K\left[l\log{\left(2\pi\epsilon\right)}+\frac{1}{2}\right].
\end{equation}
As shown in Eq.~\ref{eq:scaling_work_weak} and Eq.~\ref{eq:scaling_work_strong}, the work required for a single strong measurement at large $\epsilon$ is significantly lower than that for the concatenation of multiple weak measurements.
\section{Hierarchy of information measures}\label{appendix:hierarchy_information}
For a mixture of states $\hat\rho_\text{mixed}=\sum_x p_x \hat\rho_x$, associated with random variable $X$ of probability distribution $p_x$, the Holevo bound states that the Holevo information $\chi(X) = S[\hat\rho_\text{mixed}]-\sum_x p_x S[\hat\rho_x]$ verifies \cite{Preskill16}:
\begin{eqnarray}
    \chi(X)\geq I_{\mathbf{F}}(X:Y), \quad\forall \mathbf{F},
\end{eqnarray}
where $I_{\mathbf{F}}(X:Y)$ denotes the classical mutual information of the random variables $X$ and $Y$, the latter being generated by applying POVM $\mathbf{F} = \{\hat F_y\}$ on $\hat\rho_\text{mixed}$. More precisely, the variables $X$ and $Y$ obey, for a given choice of POVM $\mathbf{F}$,  the joint probability distribution $p_\mathbf{F}(x,y)=p_x\text{Tr}\{\hat F_y\hat\rho_x\}$, such that $I_{\mathbf{F}}(X:Y) = \sum_{x,y} p(x,y)\log\frac{p_\mathbf{F}(x,y)}{p_x p_\mathbf{F}(y)}$, with $p_\mathbf{F}(y) = \sum_x p_x p_\mathbf{F}(x,y)$.

We apply this result to the average system state after the measurement $\hat\rho_S(t_f) = \sum_r p_r \hat\rho_{S|r}$, obtained from Eq.~\eqref{eq:av_state_SA} via a partial trace over $A$, using $\hat\rho_{S|r}=\text{Tr}_A\{\hat\rho_{SA|r}\}$. We introduce random variables $r$, the measurement outcome, and $x_j$, the value of the target measurement observable. The latter is obtained from the POVM $\{\hat\Pi_j\}$, composed of the projectors onto the measurement observable eigenstates. In the main text notations, we have:
\be
\chi \geq I\left(\{j\};\{r\}\right) = \sum_{r,j} p_{r,j}\log\frac{p_{r,j}}{p_r p_j},
\ee
where we have introduced the probability distributions
\begin{eqnarray}
p_{r,j} &=& p_r\text{Tr}\{\hat \Pi_j \hat\rho_{S|r}(t_f)\}\\
p_j &=& \sum_r \text{Tr}\{\hat \Pi_j \hat\rho_{S|r}(0)\}.
\end{eqnarray}

We also note that the concavity of von Neumann entropy \cite{Preskill16}, together with the equalities $\hat\rho_S(t_f)=\sum_r p_r \hat\rho_{S|r}$ and $\hat\rho_{S|r}=\frac{1}{p_r}\sum_n p_n p(r|n) \hat\rho_{S|n}$, imply:
\be
S[\hat\rho_S(t_f)] \geq \sum_r p_r S[\hat\rho_{S|r}] \geq \sum_n p_n S[\hat\rho_{S|n}].
\ee
 Consequently:
 \be
I_q = S[\hat\rho_S(t_f)] - \sum_n p_n S[\hat\rho_{S|n}] \geq I\left(\{j\};\{r\}\right).
 \ee

 We finally have:

 \be 
S[\hat\rho_S(t_f)]\geq I_q \geq \chi(r)\geq I\left(\{j\};\{r\}\right).
 \ee

\section{Pure dephasing, energy exchange, and the conditions for SBS formation}\label{appendix:SBS_formation}
Consider an ancilla, modeled as a harmonic oscillator as in Sec.~\ref{section:qubit-cavity-model}, coupled to several subenvironments taken as independent bosonic baths, indexed by $r$ and illustrated in Fig.~\ref{fig:dephasing_different_channels}.  
The total interaction Hamiltonian reads  
\begin{equation}
    \hat{H}_I = \sum_{r} \hat{H}_{I,r}, 
    \qquad
    \hat{H}_{I,r} = \gamma_r \hat{a}^\dagger \hat{a}\otimes \hat{B}_r ,
\end{equation}
where $\gamma_r$ denotes the coupling strength and

\begin{equation}
    \hat{B}_r = \sum_{l}\hat{b}^\dagger_{r,l}\hat{b}_{r,l},
\end{equation}
is the excitation number operator in subenvironment $r$. The index $l$ runs over the many internal modes of subenvironment $r$.

Each bath is assumed to start in a thermal state, which can be expressed in the coherent-state representation as
\begin{equation}
    \hat\rho_{B,r}=\bigotimes_l \int d^2\alpha_{r,l} P_{r,l}(\alpha_{r,l})  \ket{\alpha_{r,l} }\bra{\alpha_{r,l} }
\end{equation}

with the Glauber--Sudarshan distribution  
\begin{equation}
    P(\alpha_{r,l}) = \frac{1}{\pi\bar{n}} e^{-|\alpha_{r,l}|^2/\bar{n}},
\end{equation}
and mean occupation number
\begin{equation}
    \bar{n} = \frac{1}{e^{\beta\hbar\omega}-1}.
\end{equation}

The ancilla is initially in a generic mixed state,  
\begin{equation}
    \hat{\rho}_A(0) = \sum_{n,m} \rho_A^{nm}(0) \ket{n}\bra{m},
\end{equation}
so that the total system--environment state at $t=0$ is  
\begin{equation}
    \hat{\rho}(0) = \hat{\rho}_A(0)\otimes \bigotimes_{r} \hat{\rho}_{B,r}.
\end{equation}

To describe a particular stochastic realization, one samples the initial state of each sub-environment from the distribution $P(\alpha_{r,l})$:  
\begin{equation}
    \hat{\rho}(0) = \sum_{n,m} \rho_A^{nm}(0)\, \ket{n}\bra{m}\otimes 
    \bigotimes_{r} \bigotimes_l  \ket{\alpha_{r,l}  }\bra{\alpha_{r,l} }.
\end{equation}

From this point onward, we work in the interaction picture. The corresponding unitary evolution operator factorizes as  
\begin{equation}
    \hat{U}_I(t) = e^{-i t \sum_r \hat{H}_{I,r}}
    = \sum_{n=0}^\infty \ket{n}\bra{n}\otimes\bigotimes_{r} e^{-i n \gamma_r t \hat{B}_r}.
\end{equation}
Applying this unitary to the initial state gives the evolved density operator  

\begin{equation}
    \hat{\rho}(t) = \sum_{n,m} \rho_A^{nm}(0)\, \ket{n}\bra{m}\otimes 
    \bigotimes_{r} \bigotimes_l  \ket{\alpha_{r,l} e^{-i n\gamma_{r,l} t}}\bra{\alpha_{r,l} e^{-i m\gamma_{r,l} t}}.
\end{equation}

The diagonal sector of $\hat\rho(t)$ in the ancilla eigenbasis takes the form of a SBS state
\begin{equation}
    \hat{\rho}(t) = \sum_{n} \rho_A^{nn}(0)\, \ket{n}\bra{n}\otimes 
    \bigotimes_{r} \underbrace{\bigotimes_l \ket{\alpha_{r,l} e^{-it(\omega_{r,l} + \gamma_{r,l} n)}}\bra{\alpha_{r,l} e^{-it(\omega_{r,l} + \gamma_{r,l} n)}}}_{\equiv\, \rho_r^{(n)}} + \text{off-diagonal terms},
\end{equation}
implying that the information about the ancilla state $\ket{n}$ can be obtained from a local measurement in any of the environments.

The coherences of the  ancilla density matrix decay over time due to the interaction with the sub-environments. This decay is quantified by the partial trace over the environment
\begin{equation}
    \hat{\rho}_A=\sum_{n,m}\rho_A^{nm}\prod_{r}\prod_{l}\braket{\alpha_{r,l} e^{-it\gamma_{r,l} n}|\alpha_{r,l} e^{-it\gamma_{r,l} m}},
\end{equation}
the off-diagonal terms $m\neq n$ in the previous expression are
 products of complex factors of modulus strictly smaller than $1$, whenever $\alpha_{r,l}$ is nonzero. The number of such factors in the product scales with the number of mode in each subenvironment, and the number of environment. Pure dephasing is therefore complete as long as the subenvironments are macroscopic enough. Conversely, in the diagonal terms $n=m$, the products over $r$ and $l$ are equal to unity. \\
To be able to distinguish between outcomes the trace
\begin{equation}
    \text{Tr}\{\rho_r^{(n)}\rho_r^{(m)}\} = \prod_{l} \text{Tr}\{\rho_{r,l}^{(n)}\rho_{r,l}^{(m)}\}
\end{equation}
should be much smaller than 1. Each factor in the product can be expressed in terms of the coherent-state overlap as
\begin{equation}
    \text{Tr}\{\rho_{r,l}^{(n)}\rho_{r,l}^{(m)}\}=|\braket{\alpha_{r,l} e^{-it\gamma_{r,l} n}|\alpha_{r,l} e^{-it\gamma_{r,l} m}}|^2=e^{2\alpha{r,l}^2\left(-1\cos{\left[\left(m-n\right)t\gamma_{r,l}\right]}\right)} < 1 \hspace{0.3 cm}\text{if}\hspace{0.3 cm}n\ne m,
\end{equation}
where $\alpha_{r,l}$ has been considered to be real. Since the total trace is a product over many such factors $l$, the overall overlap
$\text{Tr}\{\rho_r^{(n)}\rho_r^{(m)}\}$ becomes exponentially small as the number of modes increases.

For the same reason, the off-diagonal elements of $\hat\rho(t)$ in the ancilla energy eigenbasis decay if one trace over any single any one of the subenvironments. As a consequence, the SBS state
   \begin{align}
   \hat{\rho}(t) = \sum_{n} \rho_A^{nn}(0)\, \ket{n}\bra{n}\otimes 
    \bigotimes_{r} \rho_r^{(n)}
\end{align}
describes the statistics of any measurement done by observers having local access to the subenvironments. Detecting the coherences requires in contrast an impractical joint measurement on \emph{all} the environments.\\ 

\bibliographystyle{unsrt}
\bibliography{biblio}
\end{document}